\documentclass[11pt,double]{article}
\usepackage[latin1]{inputenc}
\usepackage{amscd}
\usepackage{amsmath}
\usepackage{amssymb}
\usepackage{amsfonts}
\usepackage{latexsym}
\usepackage{blindtext}
\usepackage{amsthm}
\usepackage{mathrsfs}
\usepackage{rotating}
\usepackage{booktabs}
\usepackage{caption}
\usepackage{xcoffins}
\usepackage[nohead]{geometry}
\usepackage{epsfig}
\usepackage{algorithm}		
\usepackage{authblk}
\usepackage{helvet}         
\usepackage{courier}        
\usepackage{url}
\usepackage{algpseudocode}
\usepackage{algpascal}
\usepackage{multirow}
\usepackage{tabularx}
\usepackage{subcaption}
\usepackage{setspace}
\usepackage[round]{natbib}
\numberwithin{equation}{section}

\newtheorem{definition}{Definition}[section]

\def\bmu{\mbox{\boldmath $\mu$}}

\def\bOmega{\mbox{\boldmath $\Omega$}}

\def\bpsi{\mbox{\boldmath $\psi$}}

\def\bnu{\mbox{\boldmath $\nu$}}

\def\bLambda{\mbox{\boldmath $\Lambda$}}

\def\bSigma{\mathbf{\Sigma}}

\def\by{\mathbf{y}} 
\def\bY{\mathbf{Y}}
\def\0{\mbox{\bf{0}}}

\def\bQ{\mathbf{Q}}

\def\sd{\mathsf{d}}

\def\sy{\mathsf{y}} 
\def\sY{\mathsf{Y}}
\def\ss{\mathsf{s}}

\def\sn{\mathsf{n}}

\def\sS{\mathsf{S}}

\def\sM{\mathsf{M}}

\newcommand{\xp}{\mathbb{E}}

\def\qmo{``}

\def\qmcsp{'' }

\newcounter{example}[section]
\def\theexample{\thesection.\arabic{example}}
{
\medskip\noindent\\
\refstepcounter{example} {\bf Example \theexample\hspace{2pt}}
-\hspace{2pt}}{%
 \begin{flushright}
 {\footnotesize $\square$}
 \end{flushright}
}

\NewCoffin\tablecoffin

\title{\large\bf Interconnected risk contributions:\\an heavy--tail approach to analyse US financial sectors}

\author[a]{Mauro Bernardi}
\author[a]{Lea Petrella\footnote{Corresponding author: Sapienza University of Rome, Dept. MEMOTEF, Via del Castro Laurenziano, 9, 00161 ROME, \textrm{Tel.:} +39.06.49766972, \textrm{email address:} \textrm{lea.petrella@uniroma1.it.}}}
\affil[a]{MEMOTEF, Sapienza University of Rome, Italy}

\begin{document}

\maketitle

\date{}

\begin{abstract}
%
%
In this paper we consider a multivariate model--based approach to measure the dynamic evolution of tail risk interdependence among US banks, financial services and insurance sectors. To deeply investigate the risk contribution of insurers we consider separately life and non--life companies. To achieve this goal we apply the multivariate student--t Markov Switching model and the Multiple--CoVaR (CoES) risk measures introduced in Bernardi \textit{et al.} \citeyearpar{bernardi_etal.2013b} to account for both the known stylised characteristics of the data and the contemporaneous joint distress events affecting financial sectors. Our empirical investigation finds that banks appear to be the major source of risk for all the remaining sectors, followed by the financial services and the insurance sectors, showing that insurance sector significantly contributes as well to the overall risk. Moreover, we find that the role of each sector in contributing to other sectors distress evolves over time accordingly to the current predominant financial condition, 
implying different interconnection strength.
%
%
\vspace{0.25cm}
\vspace{0.25cm}
\end{abstract}

\newpage

\section{Introduction}
\label{sec:intro}
%
For long time there has been a distinct separation among insurance, banking and other financial institutions in most countries, so that events in one sector usually had little effect on the others. However, during the recent years the barriers among institutions have been partly dismantled, resulting in much closer interconnections and overlaps between them and their activities. Direct and indirect dependencies among participants of a financial market contribute to the spread of malfunctions from one failing financial institution to the remaining market participants. Moreover, the latest financial crises have created a deep interest in measuring tail risk interdependence among institutions in order to evaluate the spillover effect on the real economy. The importance of the interconnectedness among institutions has been recognised to be a key requirement for the spread of a systemic crisis. In literature there exist a a variety of recent works and discussions concerning this topic (see e.g. Billio \textit{et al.}, \citeyear{billio_etal.2012}; Brechmann \textit{et al.}, \citeyear{brechmann_etal.2013}; Cont and Moussa, \citeyear{cont_moussa.2010}; Cont \textit{et al.}, \citeyear{cont_etal.2012}; Markose \textit{et al.} \citeyear{markose_etal.2012}; Podlich and Wedow, \citeyear{Podlich_etal.2013}). In line with those considerations, the aim of this paper is to investigate and fill the gap of inter--sectors analysis and to address the problem of assessing systemic importance among different sectors composing the financial system in order to examine the systemically importance financial sectors. Recently, one of the most considered co--risk measure has been the Conditional Value--at--Risk (CoVaR) of Adrian and Brunnermeier \citeyearpar{adrian_brunnermeier.2011} which measure the Value--at--Risk of an institution conditional on another institution being in financial distress. The recent literature on CoVaR  measure has been grown during the last few years (see Girardi and Erg\"un, \citeyear{girardi_ergun.2013}; Bernardi \textit{et al.}, \citeyear{bernardi_etal.2013a}; Bernardi \textit{et al.}, \citeyear{bernardi_etal.2013b}; Castro and Ferrari, \citeyear{castro_ferrari.2014};  Bernal \textit{et al.}, \citeyear{bernal_etal.2013}; Cao, \citeyear{cao.2013}; and Jaeger--Ambrozewicz, \citeyear{jaeger-ambrozewicz.2012}). However, when dealing with highly interconnected systems, it is possible that several institutions joint experience financial distress events at the same time; this situation can not be captured by the traditional CoVaR measure which consider an institution's distress per time. Underestimation of the simultaneous occurrence of interconnected rare events and the consequent bias in evaluating the transmission of risks among sectors may cause misleading policy reaction. To account for possible underestimation problems, recently Bernardi \textit{et al.} \citeyearpar{bernardi_etal.2013b} and Cao \citeyearpar{cao.2013} proposed an extension of the traditional CoVaR and Conditional Expected Shortfall (CoES), the so called Multiple--CoVaR and Mutiple--CoES, which consider joint occurrence of extreme losses. Here, following Bernardi \textit{et al.} \citeyearpar{bernardi_etal.2013b} we propose a multivariate model--based approach to measure tail risk interdependence dynamics when different sectors may experience extreme tail risk events at the same time. The approach is based on a multivariate Student--t Markov Switching model being able to capture different dynamic risk profiles through the presence of latent states, as well as to account for several stylised facts, like asymmetry, heavy tails, non linearity and persistence of extreme observations, which are crucial in financial returns time series analysis (see e.g. \citeyear{bulla.2011}). In this way we contribute to evaluate risk measures that are intrinsically dynamics since they rely on time--varying loadings of individual risk factors represented by the Value--at--Risks. The resulting dynamic evolution over time provides important monitoring tools for the market--based macro--prudential or financial stability regulation. Moreover, since we evaluate those measures on the Markov Switching predictive distribution we provide a forward--looking approach to tail risk interdependence assessment.\newline 
%
%
\indent Our analysis is mainly focused on modelling and measuring major financial sectors interconnectedness;  the idea is to stress the importance of mapping out the relationships between all those sectors highlighting the degree of their connectivities.
%
%
We achieve this goal by monitoring each sectors' total risk evolution, which may provide an early warning indicator of the global financial crisis. Moreover, measuring the contribution of each sector to the riskiness of the remaining ones, using the cross comparison of the proposed multiple risk measures,  we are able to detect the amount of interconnection among sectors.
Although most of the recent literature focuses on the banking environment, the breakdown of companies other than banks, have had a wide impact on the real economy. Sectors like insurance, who have been considered for long time safer than banks in their activities, showed a significantly altered risk profile which may affect the overall risk amount of the system (see Billio \textit{et al.}, \citeyear{billio_etal.2012}, Brechmann \textit{et al.}, \citeyear{brechmann_etal.2013} and Harrington, \citeyear{harrington.2009}). To deeply investigate the risk contribution of the insurance sector we investigate both life and non--life companies.
In particular to show the effective interdependence among financial sectors we analyse the Dow Jones US banks, the Dow Jones US financial services, the Dow Jones life and non--life insurance indexes, for the period from January 1, 1992 to June 28, 2013. To asses the relative institutions' risk contributions we calculate the Multiple--$\Delta{\rm CoVaR}$ and the Multiple--$\Delta{\rm CoES}$ which are the multiple generalization of the $\Delta{\rm CoVaR}$ and the $\Delta{\rm CoES}$ of Adrian and Brunnermeier \citeyearpar{adrian_brunnermeier.2011} in line with Bernardi \textit{et al.} \citeyearpar{bernardi_etal.2013b}. The Shapley \citeyearpar{shapley.1953} value methodology is then applied to finally attribute the overall risk shares to each institution. From an empirical point of view this is the first attempt to segment the insurance sector between life and non--life companies in a risk contribution framework. This latter aspect is quite interesting because life and non--life insurers differ by the composition of their investment portfolio which is of great importance for systemic risk management.\newline
%
\indent Our empirical findings suggest that each financial sector significantly impacts on each other during crisis periods, as well as during more stable phases. When comparing the contribution of each financial industry, banks appear to be the major source of risk for all the remaining sectors, followed by financial services and the insurance industry showing that insurance sector contribute as well to the overall risk. Moreover, we find that the role of each sector in contributing to other sectors distress evolves over time accordingly to the current predominant financial condition. We also show that banks and financial services are more interconnected than the insurance sectors and that the interconnection strengthen between banks and insurance is more evident after the 2008 global financial crisis increasing the probability of occurring joint distress events. Finally, comparing life and non--life insurance we find that they are highly interconnected both during crisis periods, as well as during financial stability phases, showing a lower interconnection after the end of the 2008 crisis.\newline
%
%
%
%
\indent The remainder of the paper is structured as follows. Section \ref{sec:hmm_models} introduces the Student--t Markov Switching model. Section \ref{sec:risk_measures} provides the definition of the risk measures used throughout the paper. Section \ref{sec:empirical_analysis} describes our empirical results while Section \ref{sec:discussion} discusses our empirical findings. Section \ref{sec:conclusion} concludes.
%
\section{The model}
\label{sec:hmm_models}
%
In this section we provide a brief description of the Markov Switching (MS) models with particular emphasis to the Student--t component distribution. The choice of this model is motivated by its attitude to well represent the financial time series dynamics capturing the underlying structure of the observations like heavy tails, asymmetry and non linear dependence. Moreover, the hidden Markov structure is able to identify periods of crisis as well as phases of financial stability. Those characteristics are particularly relevant when the main objective is to measure and manage financial or systemic risks. In particular, the MS model dynamics allows us to quantify the evolution over time and over states of the overall risk institutions' contributions. For a deeper review of MS models, see e.g. Capp\'e \textit{et al.} \citeyearpar{cappe_etal.2005}, Zucchini and MacDonald \citeyearpar{zucchini_macdonald.2009} and Dymarski \citeyearpar{dymarski.2011}. Recent applications of MS models to  financial market returns may be found in Bulla \citeyearpar{bulla.2011}, Amisano and Geweke \citeyearpar{amisano_geweke.2011}, Geweke and Amisano \citeyearpar{geweke_amisano.2010}. Recently Bernardi \textit{et al.} \citeyearpar{bernardi_etal.2013b} analysed the implications of multivariate Student--t MS models to evaluate extreme tail risk interconnectedness among financial markets participants in the bank sector. In what follows we shortly describe the model they propose to which we refer throughout the paper.\newline\newline
\noindent Let $\left\{\bY_t,t=1,\dots,T\right\}$ denote a sequence of multivariate observations, where $\bY_t = \{\sy_{1,t},\sy_{2,t},\dots,\sy_{p,t}\}\in {\mathbb R}^p$, and $\left\{\sS_t,t=1,\dots,T\right\}$ a Markov chain defined on the state space $\{1,2,\dots,L\}$. In the MS model setting 
%
the conditional distribution for the observation process $\{\bY_t\}$ depends solely on the latent state at time $t$, i.e.
\begin{equation}
f\left(\bY_t=\bY_t\mid \bY_1=\by_1,\dots,\bY_{t-1}=\by_{t-1},\sS_1=\ss_1,\dots,\sS_t=\ss_t\right)=f\left(\bY_t=\by_t\mid \sS_t=\ss_t\right),\nonumber
\end{equation}
and the unobservable process $\left\{\sS_t\right\}$ satisfies the following Markov property
\begin{equation}
\mathbb{P}\left(\sS_t=\ss_t\mid \sS_1=\ss_{1},\sS_2=\ss_{2},\dots,\sS_{t-1}=\ss_{t-1}\right)
=\mathbb{P}\left(\sS_t=\ss_t\mid \sS_{t-1}=\ss_{t-1}\right).\nonumber
\end{equation}
When dealing with financial time series it is important to account for the well known stylised facts as well as possible dependence structures among extreme events which are relevant in assessing economic risks. Those reasons motivate our assumption of multivariate Student--t distribution for modelling the observed process, i.e.
\begin{equation}
\label{eq:pdf_student}
\bY_t\mid\sS_t=\ss_t\sim\mathcal{T}_p\left(\bmu_{\ss_t},\bSigma_{\ss_t},\nu_{\ss_t}\right)
\end{equation} 
where $\left(\bmu_{l},\bSigma_{l}, \nu_{l}\right)$, $l=1,2,\dots,L$ denote location, scale and degrees of freedom parameters respectively.\newline
\indent To make inference on the unknown model parameters we consider the Expectation--Maximization (EM) approach, see Dempster \textit{et al.} \citeyearpar{dempster_etal.1977} for details while Bernardi \textit{et al.} \citeyearpar{bernardi_etal.2013b} for the algorithm under the multivariate Student--t assumption.\newline
\indent The multiple risk measures considered in the paper to asses risk interdependence are calculated on the predictive distribution of the observables in order to get a forward looking risk quantification. Let $h>0$ denotes the forecasting horizon, the predictive distribution of MS models is a finite mixture of component specific predictive distributions
\begin{eqnarray}
%
p\left(\mathbf{y}_{t+h}\mid\mathcal{I}_t\right)
=\sum_{l=1}^L\pi_{l}^{\left(h\right)}f\left(\mathbf{y}_{t+h}\mid \sS_{t+h}=l\right)
\label{eq:hmm_predictive_general}
\end{eqnarray}
with mixing weights
\begin{eqnarray}
\label{eq:predictive_mix_weights}
\pi_{l}^{\left(h\right)}
=\sum_{j=1}^L\mathbf{Q}_{j,l}^{h}\mathbb{P}\left(\sS_{t}=j\mid\mathcal{I}_t\right),
\end{eqnarray}
where $\mathcal{I}_t$ is the information up to time $t$, $\mathbf{Q}_{l,j}^{h}$ is the $\left(j,l\right)$-th entry of the Markovian transition matrix $\mathbf{Q}=\left\{q_{j,l}\right\}$ to the power $h$ with  $q_{j,l}=\mathbb{P}\left(\sS_{t}=j\mid\sS_{t-1}=l\right)$. In what follow we consider $h=1$.
%
\section{Multiple risk measures} 
\label{sec:risk_measures}
%
The main objective of the paper is to evaluate the interdependence among risky events in a multivariate setting in order to capture the simultaneous interconnectedness between insurance, banks and other financial service companies. To measure the co--movement between those institutions we refer to the Multiple--CoVaR and Multiple--CoES introduced in Bernardi \textit{et al.} \citeyearpar{bernardi_etal.2013b} who generalized the Adrian and Brunnermeier's CoVaR and CoES to account for multiple contemporaneous distress events. In this way we are able to measure the contribution of one financial sector to the riskiness of a different sector being contemporaneously linked to all the remaining ones that may themselves experience extreme losses. Moreover, as already mentioned, evaluating risk measures on MS models allows to capture the risk evolution driven by different financial and economics conditions. In what follows we define the multiple risk measures.\newline\newline 
\noindent Let $\mathcal{S}=\left\{1,2,\dots,p\right\}$ be a set of $p$ institutions, we assume that the conditioning event is a set of $d$ institutions under distress indexed by $\mathcal{J}_{\sd}=\left\{j_1,j_2,\dots,j_d\right\}\subset\,_d\mathbb{C}_{p-1}$, where $_d\mathbb{C}_{p-1}$ is the set of all possible combinations of $p-1$ elements of class $d$, with $d\leq p-1$. Moreover, assuming that institution $i\in\mathcal{S}$ with $i\notin\mathcal{J}_{\sd}$ and $\mathcal{J}_{\sn}=\overline{\mathcal{J}_\sd}$ is the set of institutions being in the \qmo normal\qmcsp state, we define the ``\textit{Multiple}--CoVaR'', $\text{CoVaR}_{i\vert\mathcal{J}_\sd}^{\tau_1\vert \tau_2}$ as follows.
\begin{definition} 
\label{def:covar_multiple}
Let $\bY=\left(\sY_1,\sY_2,\dots,\sY_i,\dots,\sY_p\right)$ be the vector of institution returns, then ${\rm CoVaR}_{i\vert\mathcal{J}_\sd}^{\tau_1\vert \tau_2}$ is the Value--at--Risk of institution $i\in\mathcal{S}$, conditional on the set of institutions $\mathcal{J}_{\sd}$ being at their individual ${\rm VaR}_{\tau_2}$--level $\hat{\by}_{\mathcal{J}_\sd}^{\tau_2}=\left(\hat{\sy}_{j_1}^{\tau_2},\hat{\sy}_{j_2}^{\tau_2},\dots,\hat{\sy}_{j_d}^{\tau_2}\right)$ and the set of institutions $\mathcal{J}_{\sn}=\overline{\mathcal{J}_\sd}$ being at their individual ${\rm VaR}_{0.5}$--level $\hat{\by}_{\mathcal{J}_\sn}^{0.5}=\left(\hat{\sy}_{j_{d+1}}^{0.5},\hat{\sy}_{j_{d+2}}^{0.5},\dots,\hat{\sy}_{j_{p-1}}^{0.5}\right)$
i.e., ${\rm CoVaR}_{i\vert\mathcal{J}_\sd}^{\tau_1\vert \tau_2}$ satisfies the following equation
\begin{eqnarray}
\mathbb{P}\left(\sY_i\leq {\rm CoVaR}_{i\vert\mathcal{J}_\sd}^{\tau_1\vert \tau_2}
\,\mid \bY_{\mathcal{J}_\sd}=\hat{\by}_{\mathcal{J}_\sd}^{\tau_2},\bY_{\mathcal{J}_\sn}=\hat{\by}_{\mathcal{J}_\sn}^{0.5}\right)=\tau_1,\quad\forall i=1,2,\dots,p. 
\label{eq:mcovar_def}
\end{eqnarray}
%
%
\end{definition}
\noindent The lack of subadditivity property of the Value--a--Risk suggests to introduce, in addition to the CoVaR, the Conditional Expected Shortfall (CoES), defined by Adrian and Brunnermeier \citeyearpar{adrian_brunnermeier.2011} as the Expected Shortfall (ES) evaluated on the conditional distribution of $\sY_i$ given $\sY_j$, for two different institutions $i$ and $j$. The following definition characterises the extension of CoES to the Multiple--CoES accounting for multiple contemporaneous distress events.
\begin{definition}
Let $\bY=\left(\sY_1,\sY_2,\dots,\sY_p\right)$ be the vector of institution returns, then the ${\rm CoES}_{i\vert\mathcal{J}_\sd}^{\tau_1\vert \tau_2}$
is the Expected Shortfall of institution $i\in\mathcal{S}$, conditional on the set of institutions $\mathcal{J}_\sd$ being at their individual ${\rm ES}_{\tau_2}$-level $\widehat{\bpsi}_{\by_{\mathcal{J}_\sd}}\left(\hat{\by}_{\mathcal{J}_\sd}^{\tau_2}\right)=\left(\hat{\psi}_{\sy_{j_1}}\left(\hat{\sy}_{j_1}^{\tau_2}\right),\hat{\psi}_{\sy_{j_2}}\left(\hat{\sy}_{j_2}^{\tau_2}\right),\dots,\hat{\psi}_{\sy_{j_d}}\left(\hat{\sy}_{j_d}^{\tau_2}\right)\right)$ and the set of institutions $\mathcal{J}_\sn$ being at their individual ${\rm ES}_{0.5}$--level $\widehat{\bpsi}_{\by_{\mathcal{J}_\sn}}\left(\hat{\by}_{\mathcal{J}_\sn}^{\tau_2}\right)=\left(\hat{\psi}_{\sy_{j_{d+1}}}\left(\hat{\sy}_{j_{d+1}}^{\tau_2}\right),\hat{\psi}_{\sy_{j_{d+2}}}\left(\hat{\sy}_{j_{d+2}}^{\tau_2}\right),\dots,\hat{\psi}_{\sy_{j_{p-1}}}\left(\hat{\sy}_{j_{p-1}}^{\tau_2}\right)\right)$, with $\hat{\psi}_{\sy_{j}}\left(\hat{\sy}_{j}^{\tau}\right)\equiv{\rm ES}_{\tau}\left(\sY_j\right)$, $\forall j=1,2,\dots,d$, and can be defined in the following way
\begin{eqnarray}
{\rm CoES}_{i\vert\mathcal{J}_\sd}^{\tau_1\vert \tau_2}\equiv\xp\left(\sY_i\mid\sY_i\leq\hat{\sy}_i^{\tau_1},\bY_{\mathcal{J}_\sd}=\widehat{\bpsi}_{\by_{\mathcal{J}_\sd}}\left(\hat{\by}_{\mathcal{J}_\sd}^{\tau_2}\right),\bY_{\mathcal{J}_\sn}=\widehat{\bpsi}_{\by_{\mathcal{J}_\sn}}\left(\hat{\by}_{\mathcal{J}_\sn}^{0.5}\right)\right).
\label{eq:mcoes_def}
\end{eqnarray}
\end{definition}
\noindent Analytical formulae for the Multiple--CoVaR and Multiple--CoES defined in equation \eqref{eq:mcovar_def} and \eqref{eq:mcoes_def} respectively, under the Multivariate Gaussian and Student--t assumption are provided in Bernardi \textit{et al.} \citeyearpar{bernardi_etal.2013b}.\newline
\indent In order to quantify the marginal contribution of individual institutions, we consider the Multiple--$\Delta$CoVaR ($\Delta^\sM\text{CoVaR}$) and the  Multiple--$\Delta$CoES ($\Delta^\sM\text{CoES}$) as straightforward generalization of the $\Delta$CoVaR and $\Delta$CoES of Adrian and Brunnermaier defined as follows: 
%
\begin{eqnarray}
\Delta^\sM{\rm CoVaR}_{i\vert\mathcal{J}_\sd}^{\tau_1\vert \tau_2}&=&
{\rm CoVaR}_{\tau_1}\left(\sY_i\mid\bY_{\mathcal{J}_\sd}=\hat{\by}_{\mathcal{J}_\sd}^{\tau_2},\bY_{\mathcal{J}_\sn}=\hat{\by}_{\mathcal{J}_\sn}^{0.5}\right)-\nonumber\\
&&\qquad\qquad\qquad\qquad{\rm CoVaR}_{\tau_1}\left(\sY_i\mid\bY_{\mathcal{J}_\sd\cup\mathcal{J}_\sn}=\hat{\by}_{\mathcal{J}_\sd\cup\mathcal{J}_\sn}^{0.5}\right)
\label{eq:delta_m_covar}
\end{eqnarray}
\begin{eqnarray}
\Delta^\sM{\rm CoES}_{i\vert\mathcal{J}_\sd}^{\tau_1\vert \tau_2}&=&
{\rm ES}_{\tau_1}\left(\sY_i\mid\bY_{\mathcal{J}_\sd}=\hat{\by}_{\mathcal{J}_\sd}^{\tau_2},\bY_{\mathcal{J}_\sn}=\hat{\by}_{\mathcal{J}_\sn}^{50\%}\right)-\nonumber\\
&&\qquad\qquad\qquad\qquad{\rm ES}_{\tau_1}\left(\sY_i\mid\bY_{\mathcal{J}_\sd\cup\mathcal{J}_\sn}=\hat{\by}_{\mathcal{J}_\sd\cup\mathcal{J}_\sn}^{50\%}\right).
\label{eq:delta_m_es}
\end{eqnarray}
%
%
\noindent Since different sets of institutions belonging to the conditioning distress events qualify different measures of risk contribution, to compose the puzzle of overall risk attribution to each institution we apply the Shapley value methodology initially proposed by Shapley \citeyearpar{shapley.1953} in the field of cooperative games. The idea of applying the Shapley value methodology to the systemic risk attribution has been previously considered by Tarashev \textit{et al.} \citeyearpar{tarashev_etal.2010}, Cao \citeyearpar{cao.2013} and Bernardi \textit{et al.} \citeyearpar{bernardi_etal.2013b}. The portion of the overall value that the Shapley methodology attributes to each of the players in a cooperative game equals the average of this player's marginal contribution to the value created by all possible permutations on the set of players as evaluated by the $\Delta^\sM$CoVaR and $\Delta^\sM$CoES risk measures. In our setting this value coincides with the overall risk generated by the market participants. The additivity axiom satisfied by the Shapley value methodology ensures that the risk allocation is efficient in the sense that the shares of overall risk attributed to individual institutions exactly sum to the total risk, i.e the $\Delta^\sM$CoVaR (or $\Delta^\sM$CoES) of all the financial institutions in the system being in distress. 
%
%
%
\section{Empirical Analysis}
\label{sec:empirical_analysis}
%
In this section, we focus on the interconnectedness between banking, financial services and insurance sectors during the period 1992--2013 with particular emphasis on the distinction between life and non--life insurance companies. life and non--life insurers mainly differ for their exposure to risk as well as for the composition of their respective investment portfolios which is quite relevant for systemic risk management and it is strictly connected with the core business of the company. Concerning the first aspect, life insurance companies are more exposed than non--life ones to the risk related to catastrophic mortality events. Moreover, life insurance companies usually have longer investment horizons and this characteristic largely contributes to increase their riskiness with respect to events which affect aggregate financial market downturns or contractions in the aggregate income. On these basis we argue that financial crisis should impact more the life insurance industry than the non--life one. Concerning their respective individual contribution to the overall risk it is of relevance, for risk policy purposes, to assess whether they actively contribute to spread risks over different sectors or if they instead are victims of negative contagion effects. To examine how the overall risk shares among the US banking, insurance and the more general financial sectors, we use the multivariate Student--t MS model and the methodology explained in the previous sections to four sector indexes belonging to the Dow Jones Industrial Average (DJIA) index. The next section describes the data used in our empirical analysis while sections \ref{sec:results} and \ref{sec:systemic_risk} detail our main findings.
%
%
%
%
%
%
%
%
\subsection{The data}
\label{sec:data}
%
%
\begin{figure}[!t]
\begin{center}
\captionsetup{font={footnotesize}, labelfont=sc}
\includegraphics[width=0.90\linewidth, height=0.3\linewidth]{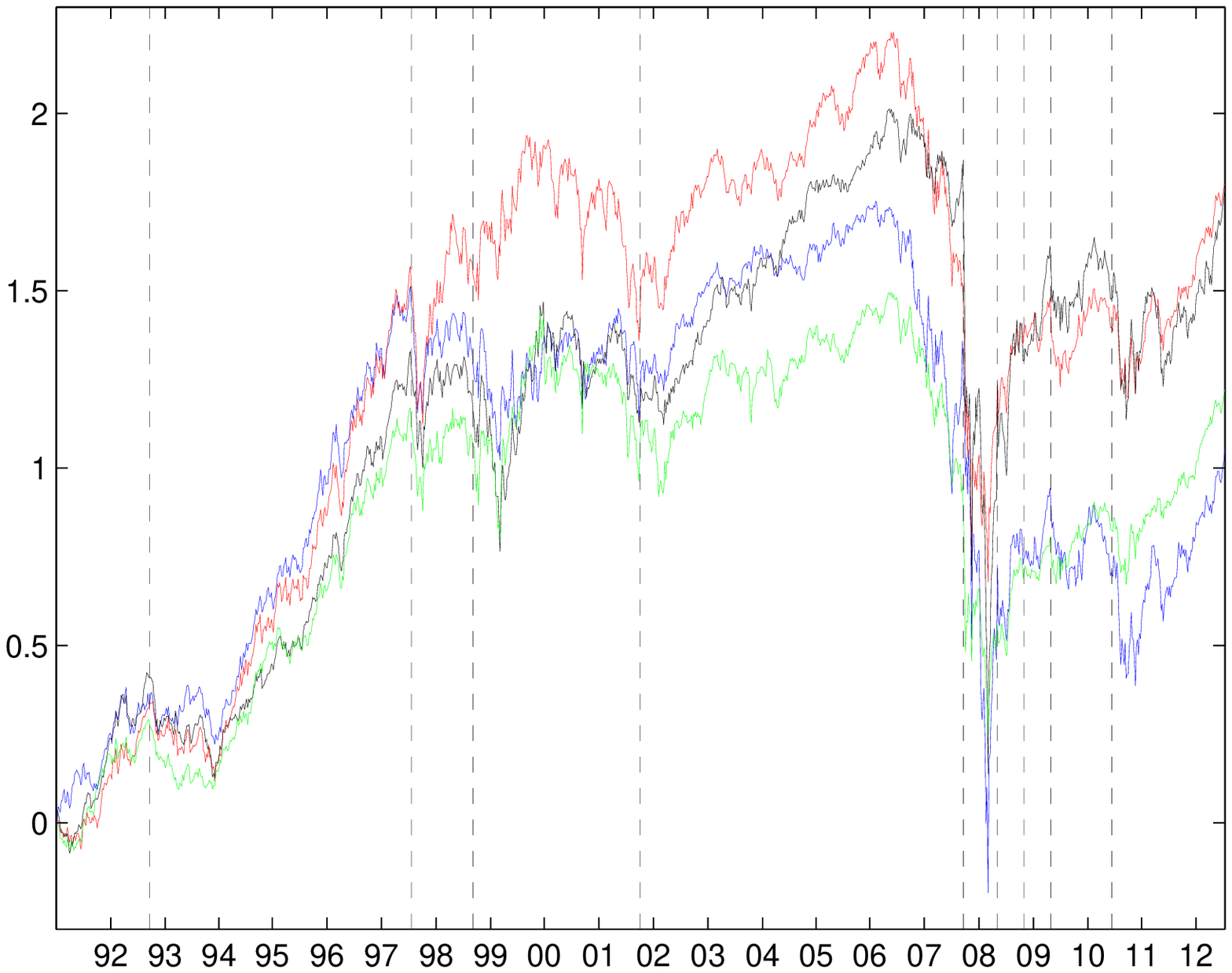}
\caption{\footnotesize{Cumulative returns of the different sectors: banks \textit{(blue line)}, financial services \textit{(red line)}, life insurance \textit{(dark line)} and non--life insurance \textit{(green line)}.\newline\newline
Vertical dotted lines represent major financial downturns: the ``Black Wednesday'' (September 16, 1992), the Asian crisis (July, 1997), the Russian crisis (August, 1998), the September 11, 2001 shock, the Bear Stearns hedge funds collapse (August 5, 2007), the Bear Stearns acquisition by JP Morgan Chase, (March 16, 2008), the Lehman's failure (September 15, 2008), the peak of the onset of the recent global financial crisis (March 9, 2009) and the European sovereign-debt crisis of April 2010 (April 23, 2010, Greek crisis).}}
\label{fig:USSectors_cumret}
\end{center}
\end{figure}
%
To assess the overall risk contribution of life and non-life insurances, banks and other financial institution in the US market, we consider four sector indexes belonging to the US Dow Jones Composite Index: banks, financial services, life and non--life insurance. Market weekly returns of the four sector's indexes spans the period from January 1, 1992 to June 28, 2013. Full sample descriptive statistics are provided in Table \ref{tab:USSectors_data_summary_stat} in Appendix \ref{sec:appendix_A}. We observe that both insurance sectors are characterised by a more pronounced skewness with respect to the banks and diversified financial services sectors. Surprisingly, between the two insurance sectors, the life one displays the largest kurtosis index, a value in line with that observed for banks and significantly different from the one observed for non--life. In addition, the Jarque--Bera (JB) statistic confirms the departure from normality for all return series at the 1\% level of significance. Moreover, the 1\% empirical quantile in column eight of Table \ref{tab:USSectors_data_summary_stat} supports the idea that life insurance individually considered would be the riskiest sector among those considered in this analysis. However, since our main concern is to investigate how sources of risk spread among different sectors, we need to gather informations about their joint dynamic evolution. In an unreported analysis, as a first step we evaluate the full sample correlation between sectors noting that non--life insurance is less correlated to the banks sector than the life one and displays the largest coefficient with the diversified financial service sector. On these grounds one should argues that the market co--movements between banks and life insurers should be larger than intra--sectorial ones (the two kind of insurers we consider here). In Section \ref{sec:results} we show that this evidence can be misleading when considering extreme interdependent events. One possible explanation for this discordant results can be ascribed to the presence of non--linear relations among asset returns otherwise captured by assuming a Student--t distribution for the conditional density of the MS model.\newline
\indent Figure \ref{fig:USSectors_cumret} shows the time series of cumulative returns for all the considered assets, from January 2nd, 1992 till the end of the sample. Vertical dotted lines refer to the following events: the ``Black Wednesday'' (September 16, 1992), the Asian crisis (July, 1997), the Russian crisis (August, 1998), the September 11--2001 shock, the onset of the mortgage subprime crisis identified by the Bear Stearns hedge funds collapse (August 5, 2007), the Bear Stearns acquisition by JP Morgan Chase, (March 16, 2008), the Lehman's failure (September 15, 2008), the peak of the onset of the recent global financial crisis (March 9, 2009) and the European sovereign--debt crisis of April 2010 (April 23, 2010, Greek crisis). The figure gives insights about how the crisis periods affect the different sectors here considered. After the 2001 Twin Towers attack till the middle of 2007, the US financial system experienced a long period of small perturbations and stability ended shortly after the collapse of two Bear Stearns hedge funds in early August 2007. Starting from August 2007, the financial market experiences a huge fall, the subprime mortgage crisis that led to a financial crisis and subsequent recession that began in 2008. Several major financial institutions collapsed in September 2008, with significant disruption in the flow of credit to businesses and consumers and the onset of a severe global recession. The system hit the bottom in March 2009, and then started a slow recovery which culminated just before the European sovereign--debt crisis of April 2010. It is interesting to note that, since the beginning of the 2007 global crisis all the considered sectors experienced huge capital losses with the banking sector (blue line) being the most affected by the crisis. Moreover, banks and non--life insurance, on one hand, and financial services and life insurance, on the other hand, become more related after European sovereign--debt crisis, displaying similar trends.
%
\subsection{Estimation Results}
\label{sec:results}
%
In this section we proceed by fitting the best model for the data considered and by estimating all the parameters involved. 
To select the best model in the multivariate Student--t MS setting we need to chose the number $L$ of latent states  $\left\{\sS_t,t=1,\dots,T\right\}$. According to the current literature (see e.g. Capp\'e \textit{et al.} \citeyear{cappe_etal.2005}; Ryd\'en, \citeyear{ryden.2008}), we apply the Akaike Information Criterion (AIC) and the Bayesian Information Criterion (BIC) which involve different penalisation terms depending on the number of non--redundant parameters. In particular, we fit the proposed model with a number of hidden states $L$ from 2 to 6. The results of this preliminary analysis are reported in Table \ref{tab:DataSet_Model_Selection} in Appendix \ref{sec:appendix_A} where it is evident that both information criteria prefer the model with four hidden states.\newline
\indent For the selected model, Tables \ref{tab:DataSet_International_Estimates_2} and \ref{tab:DataSet_International_Estimates_1} in Appendix \ref{sec:appendix_A} summarise parameter estimates. The dynamical evolution of risk--return profiles, often documented in the financial literature, is well captured by our model. In fact, we observe in Table \ref{tab:DataSet_International_Estimates_1}, that two positive (state 1 and 2) and two negative (state 3 and 4) regimes are identified according to significantly different state--specific return means. Furthermore, large negative returns (state 1) are characterised by quite large standard deviations (parameter $\Lambda$) as opposed to negative returns where standard deviations are substantially lower. States 2 and 3 identify periods of low volatility associated with moderately negative and positive mean returns respectively. This essentially implies that state 1 and 4 can be identified as periods of financial turbulence and stability, while state 2 and 3 are regimes where the financial system transits just after or immediately before a crisis periods. This latter observation can be evinced also by inspecting Figure \ref{fig:USSectors_SmoothProb} displaying the Markovian smoothed probabilities of being in a given state at each time period $\mathbb{P}\left(\sS_t\mid\mathcal{I}_T\right)$. During the 2007--2008 global financial crisis, for example, we observe that the probability of being in state 1 (turbulence) is larger than 99\%. During the period immediately before the 2007 crisis, covering most of the 2006 and 2007 years, till the Bearn Stearns hedge fund collapse of August 2007, the system visits the transitory state 2, which correspond to the pre--crisis regime. All those results document the importance of choosing the right model specification in order to understand the global dynamics of the economic system.\newline 
\indent As extensively documented in Bernardi \textit{et al.} \citeyearpar{bernardi_etal.2013b}, the multivariate Student--t approach considered here, is also able to identify different contagion effects among stocks, measured by the state--specific correlations (parameter $\Omega$). As expected, correlations are higher during crisis period while, during more stable phases, variances are relatively low and the contagion effect is less marked.\newline
\indent Table \ref{tab:DataSet_International_Estimates_2} provides parameter estimates of the Student--t degrees--of--freedom $\boldsymbol{\nu}$ and the transition probabilities $\bQ$ of the hidden Markov chain. Looking at the $\bnu$ parameters estimate fat--tails have been detected and this is in line with empirically observed stylised facts. Moreover, the introduction of conditional Student--t distributions increases the state persistence significantly, resulting in longer and more stable volatility periods. This is confirmed by the transition matrix estimate. The large off--diagonal transition probabilities in all states except the first one, confirms the large persistence of the transitory states (2 and 3) as well as the state 4 of financial stability. State 1 of financial crisis is instead characterised by a smaller probability value on the main diagonal and by a moderately large probability to move to state 2, suggesting that after coming out from a crisis the system enters a period of \qmo moderate\qmcsp financial turbulence.\newline
\indent It important to note that, although parameter estimates provide relevant information to support the policy decision making process, they do not provide enough insights to evaluate extreme tail interdependence.
%
\subsection{Overall risk contributions}
\label{sec:systemic_risk}
%
\begin{figure}[!t]
\begin{center}
\captionsetup{font={footnotesize}, labelfont=sc}
\includegraphics[width=0.9\linewidth, height=0.4\linewidth]{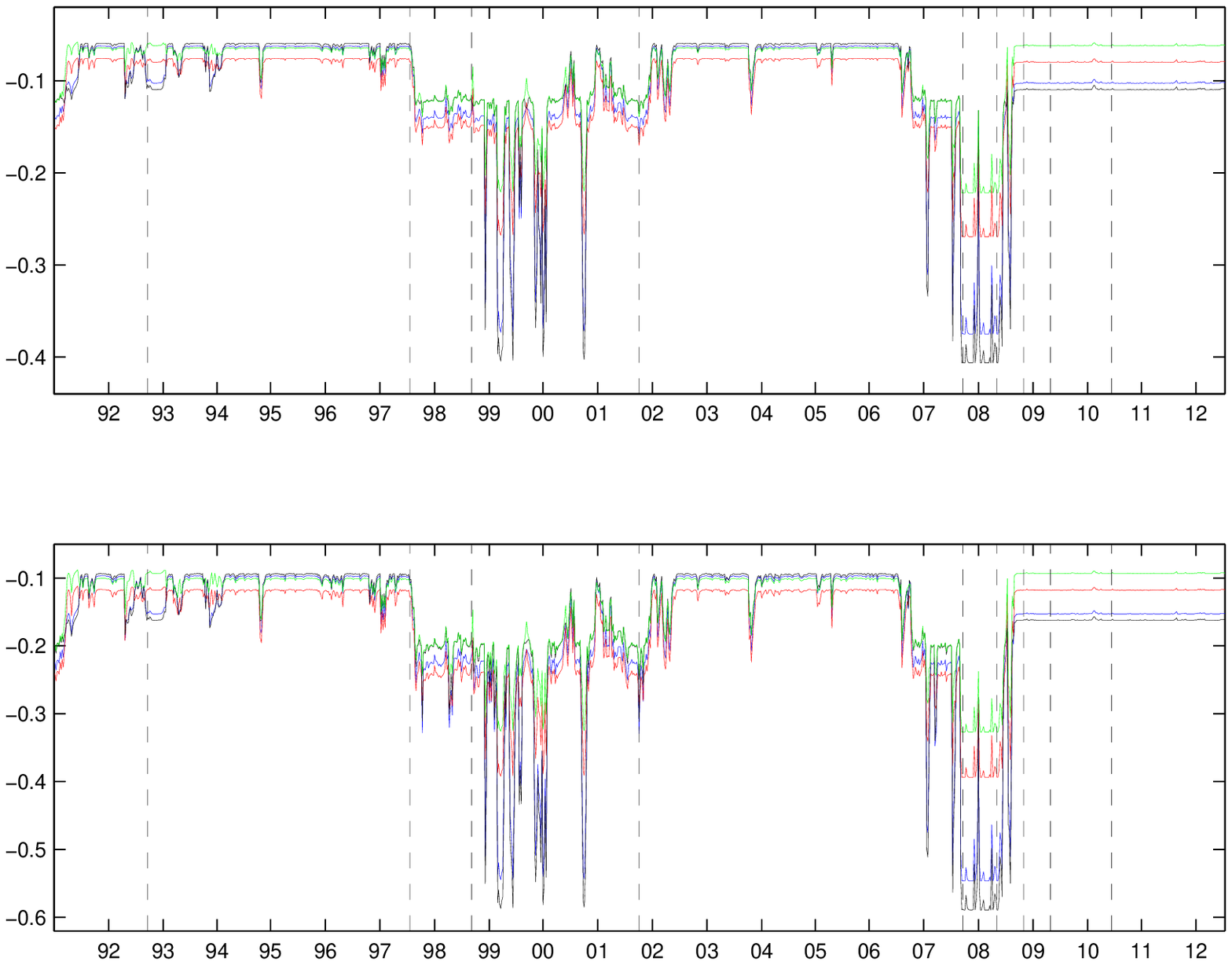}
\caption{\footnotesize{Total risk evaluated by Multiple--CoVaR \textit{(Top panel)} and Multiple--CoES \textit{(Bottom panel)} for the different sectors: banks \textit{(blue line)}, financial services (red line), life insurance \textit{(dark line)} and non--life insurance \textit{(green line)}.
%
}}
\label{fig:USSectors_Total_Risk}
\includegraphics[width=0.9\linewidth, height=0.4\linewidth]{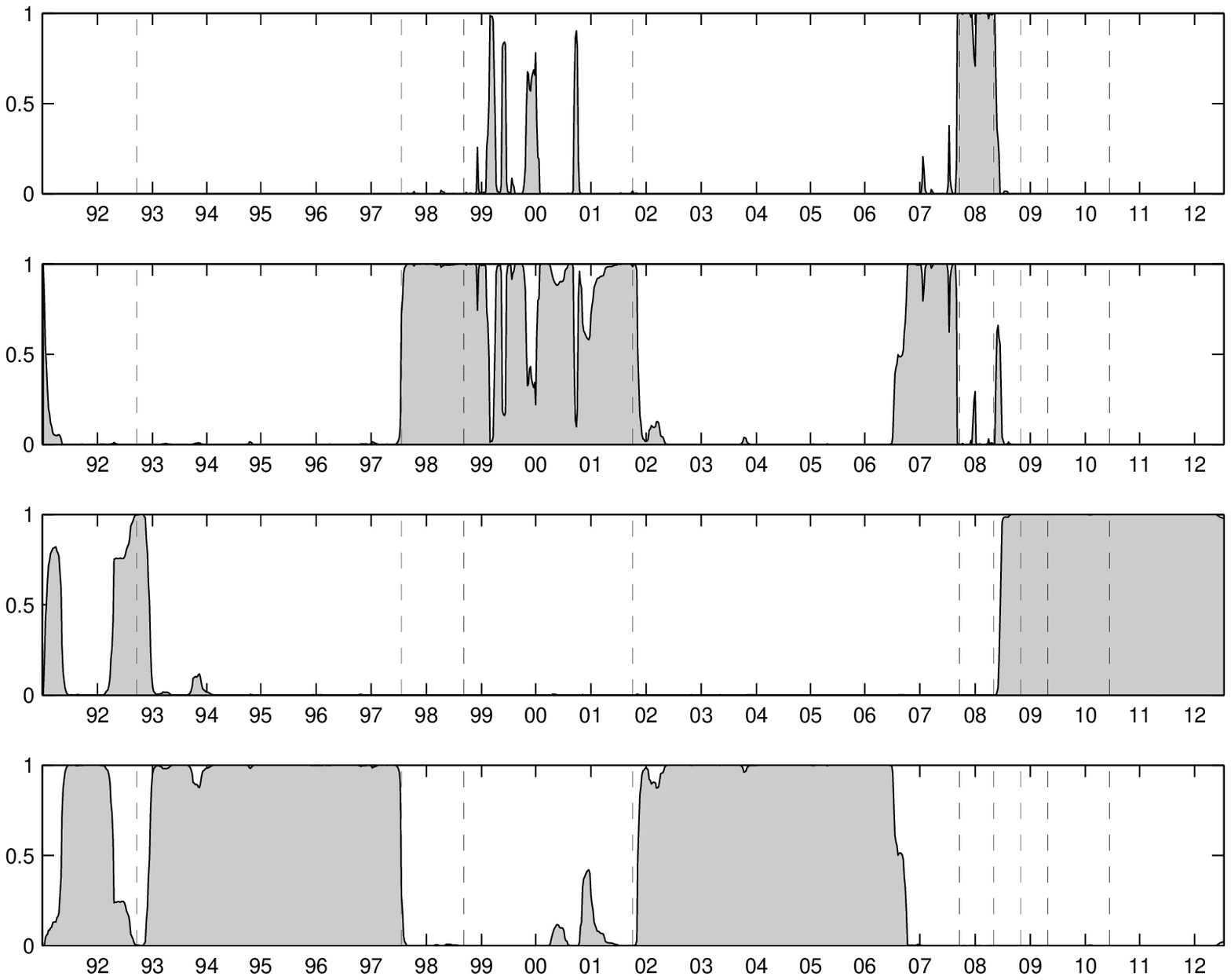}
\caption{\footnotesize{Smoothed probabilities of visiting the states of the Markovian chain $\mathbb{P}\left(\sS_t=j\mid\mathcal{I}_T\right)$ for $j=1,\dots,4$ (from top to bottom) implied by the Student--t model with four components.\newline\newline
Vertical dotted lines represent major financial downturns: the ``Black Wednesday'' (September 16, 1992), the Asian crisis (July, 1997), the Russian crisis (August, 1998), the September 11, 2001 shock, the Bear Stearns hedge funds collapse (August 5, 2007), the Bear Stearns acquisition by JP Morgan Chase, (March 16, 2008), the Lehman's failure (September 15, 2008), the peak of the onset of the recent global financial crisis (March 9, 2009) and the European sovereign-debt crisis of April 2010 (April 23, 2010, Greek crisis).}}
\label{fig:USSectors_SmoothProb}
\end{center}
\end{figure}
%

%
\begin{figure}[!t]
\begin{center}
\begin{subfigure}[b]{0.3\textwidth}
\includegraphics[width=1.0\linewidth, height=0.6\linewidth]{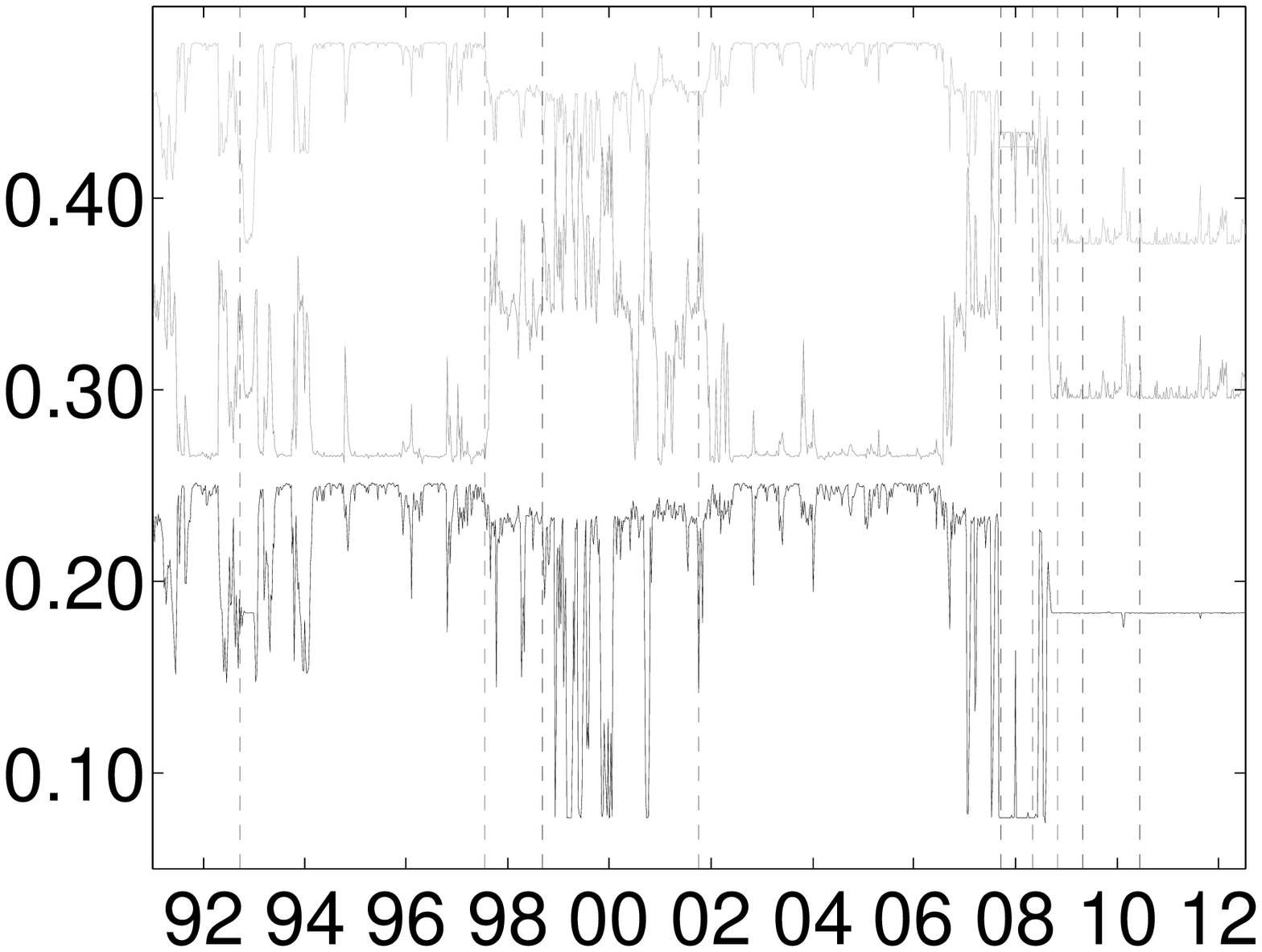}
\caption{\footnotesize{Financial services \textit{(light gray)}, life \textit{(gray)} and non--life \textit{(dark)} insurance against banks.}}
\label{fig:SV_CoVaR_banks}
\end{subfigure}\qquad\qquad
%
%
\begin{subfigure}[b]{0.3\textwidth}
\includegraphics[width=1.0\linewidth, height=0.6\linewidth]{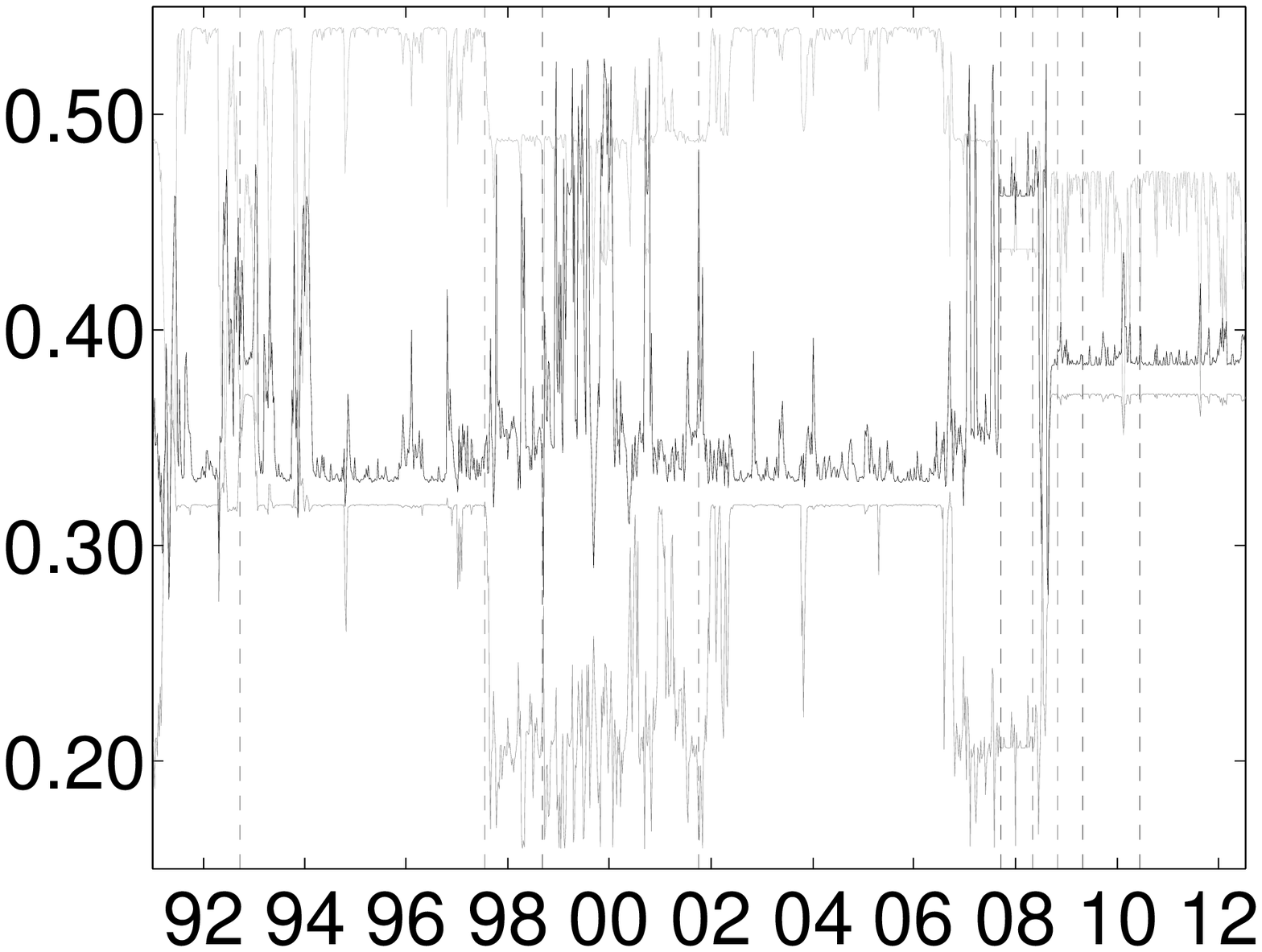}
\caption{\footnotesize{Banks \textit{(light gray)}, life \textit{(gray)} and non--life \textit{(dark)} insurance against financial services.}}
\label{fig:SV_CoVaR_financials}
\end{subfigure}\\
\vspace{1.0cm}
\begin{subfigure}[b]{0.3\textwidth}
\includegraphics[width=1.0\linewidth, height=0.6\linewidth]{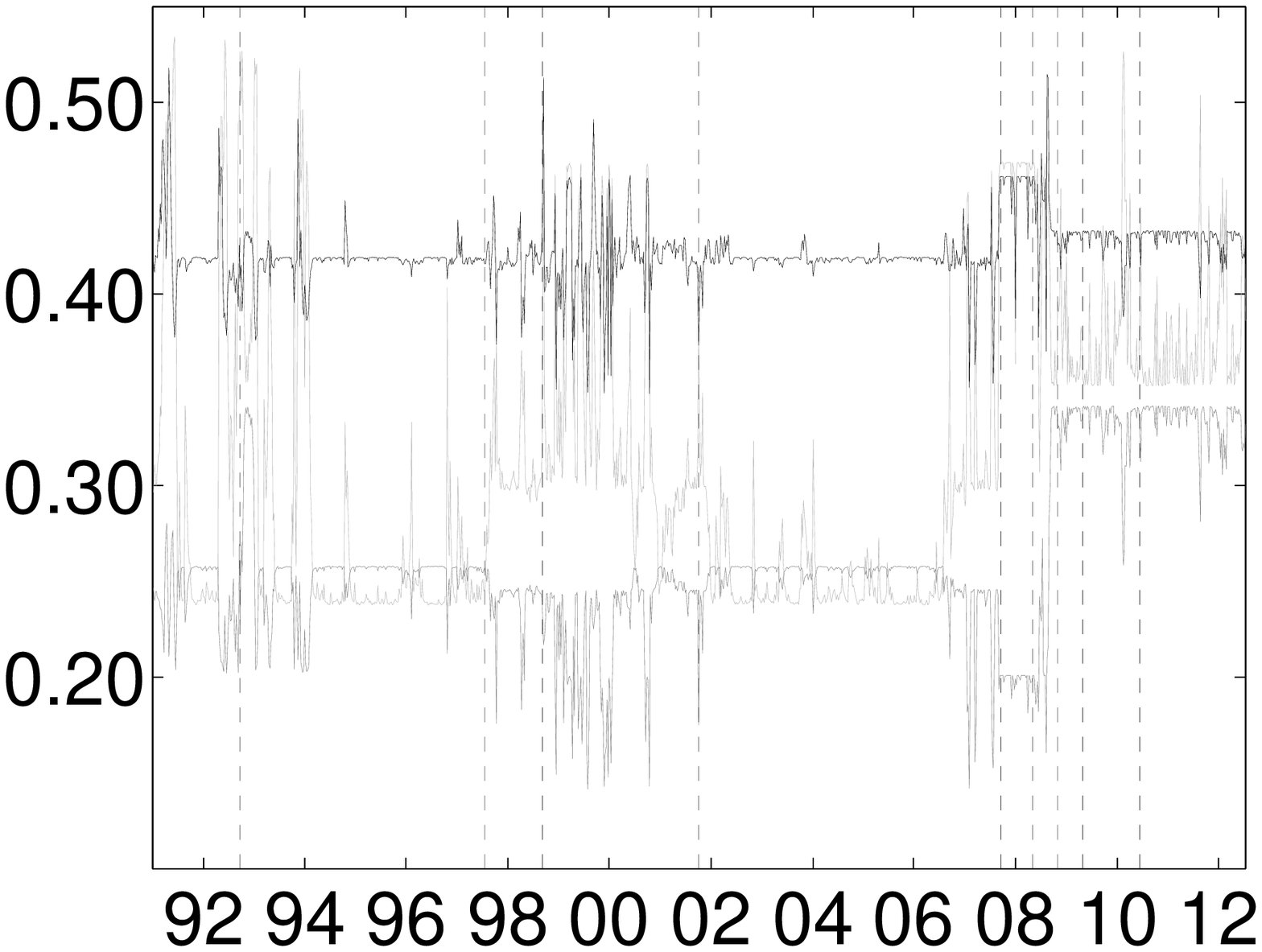}
\caption{\footnotesize{Banks \textit{(light gray)}, financial services \textit{(gray)} and non--life insurance \textit{(dark)} against life insurance.}}
\label{fig:SV_CoVaR_life}
\end{subfigure}\qquad\qquad
%
%
\begin{subfigure}[b]{0.3\textwidth}
\includegraphics[width=1.0\linewidth, height=0.6\linewidth]{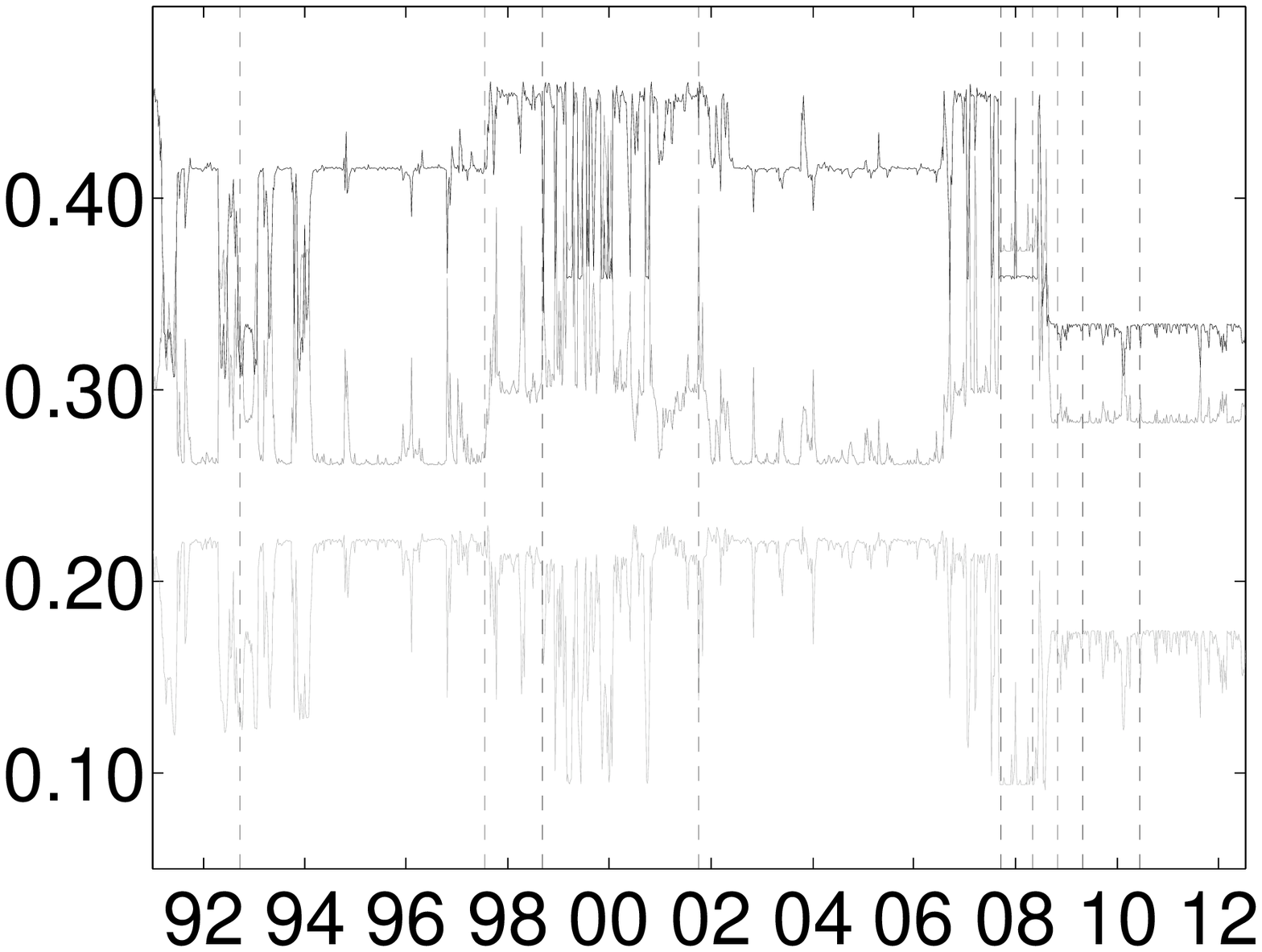}
\caption{\footnotesize{Banks \textit{(light gray)}, financial services \textit{(gray)} and life insurance \textit{(dark)} against non--life insurance.}}
\label{fig:SV_CoVaR_non-life}
\end{subfigure}
\captionsetup{font={footnotesize}, labelfont=sc}
\caption{\footnotesize{Shalpey value based on $\Delta^\sM$--CoVaR of the different sectors against banks (\subref{fig:SV_CoVaR_banks}), financial services (\subref{fig:SV_CoVaR_financials}), life insurance (\subref{fig:SV_CoVaR_life}) and non--life insurance (\subref{fig:SV_CoVaR_non-life}).
Vertical dotted lines represent major financial downturns: the ``Black Wednesday'' (September 16, 1992), the Asian crisis (July, 1997), the Russian crisis (August, 1998), the September 11, 2001 shock, the Bear Stearns hedge funds collapse (August 5, 2007), the Bear Stearns acquisition by JP Morgan Chase, (March 16, 2008), the Lehman's failure (September 15, 2008), the peak of the onset of the recent global financial crisis (March 9, 2009) and the European sovereign-debt crisis of April 2010 (April 23, 2010, Greek crisis).}}
\label{fig:USSectors_Shapley_value_CoVaR_ALL}
\end{center}
\end{figure}
%

%
\begin{figure}[!t]
\begin{center}
\begin{subfigure}[b]{0.3\textwidth}
\includegraphics[width=1.0\linewidth, height=0.6\linewidth]{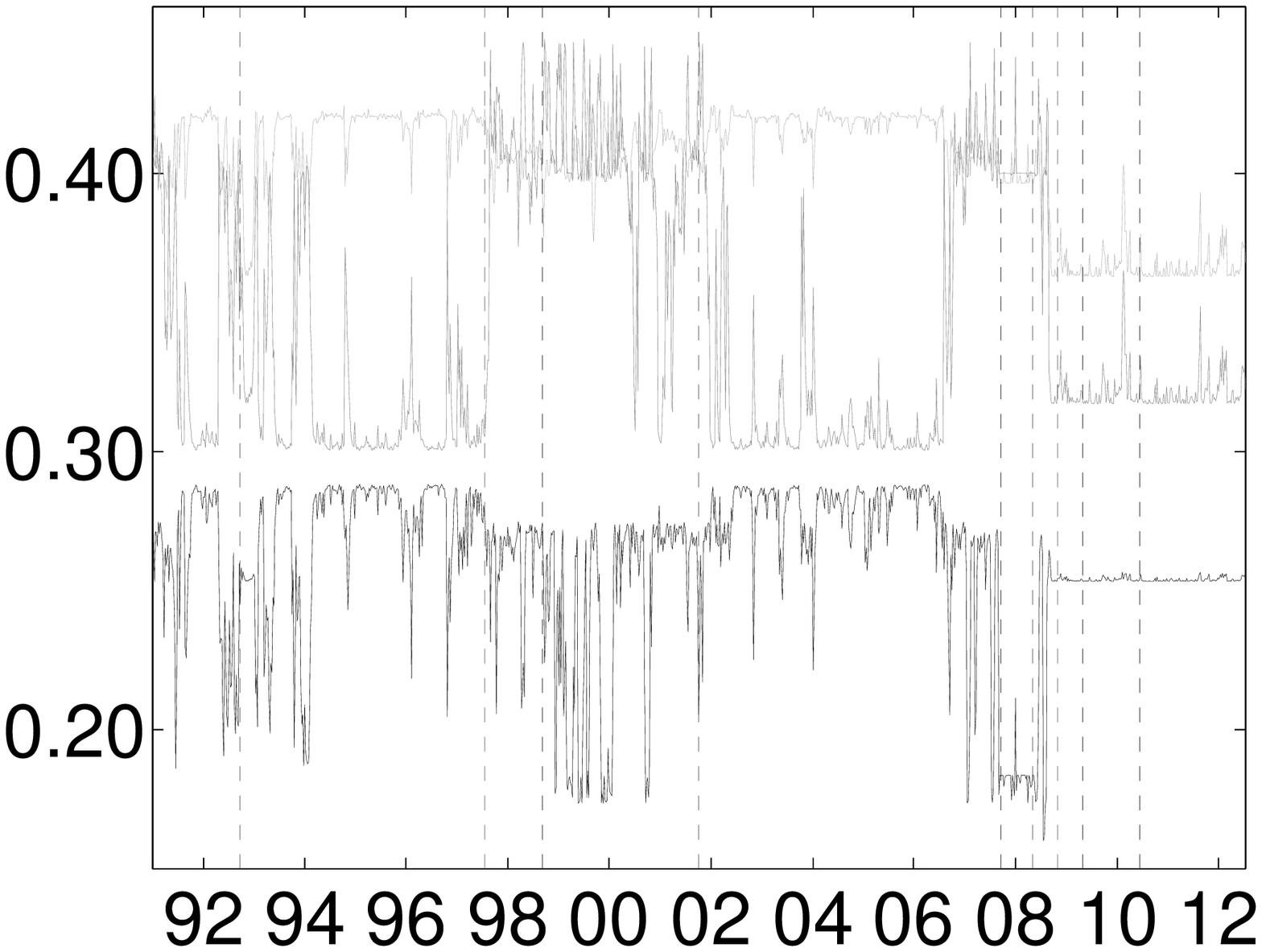}
\caption{\footnotesize{Financial services \textit{(light gray)}, life \textit{(gray)} and non--life \textit{(dark)} insurance against banks.}}
\label{fig:SV_CoES_banks}
\end{subfigure}\qquad\qquad
\begin{subfigure}[b]{0.3\textwidth}
\includegraphics[width=1.0\linewidth, height=0.6\linewidth]{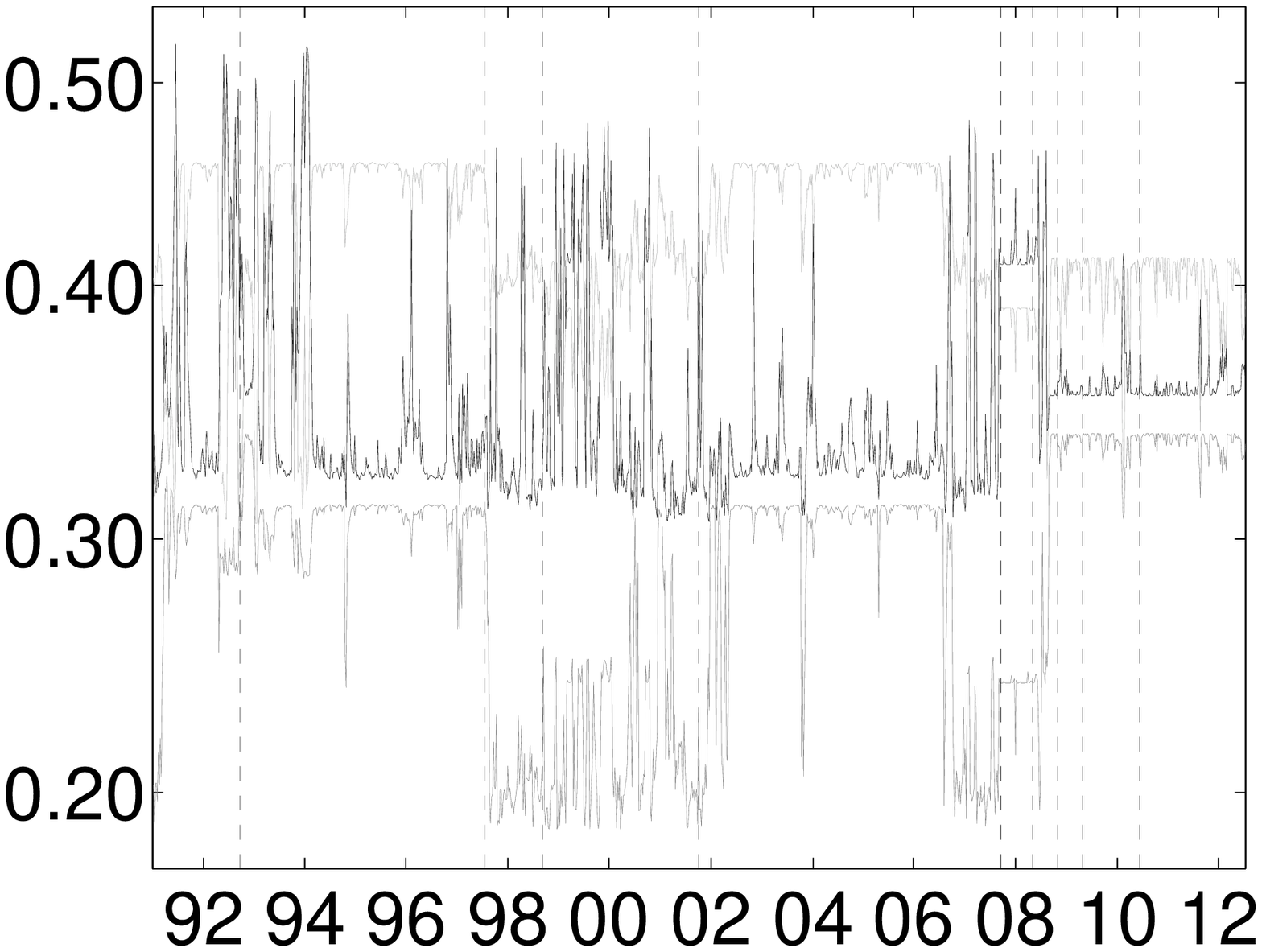}
\caption{\footnotesize{Banks \textit{(light gray)}, life \textit{(gray)} and non--life \textit{(dark)} insurance against financial services.}}
\label{fig:SV_CoES_financials}
\end{subfigure}\\
\vspace{1.0cm}
\begin{subfigure}[b]{0.3\textwidth}
\includegraphics[width=1.0\linewidth, height=0.6\linewidth]{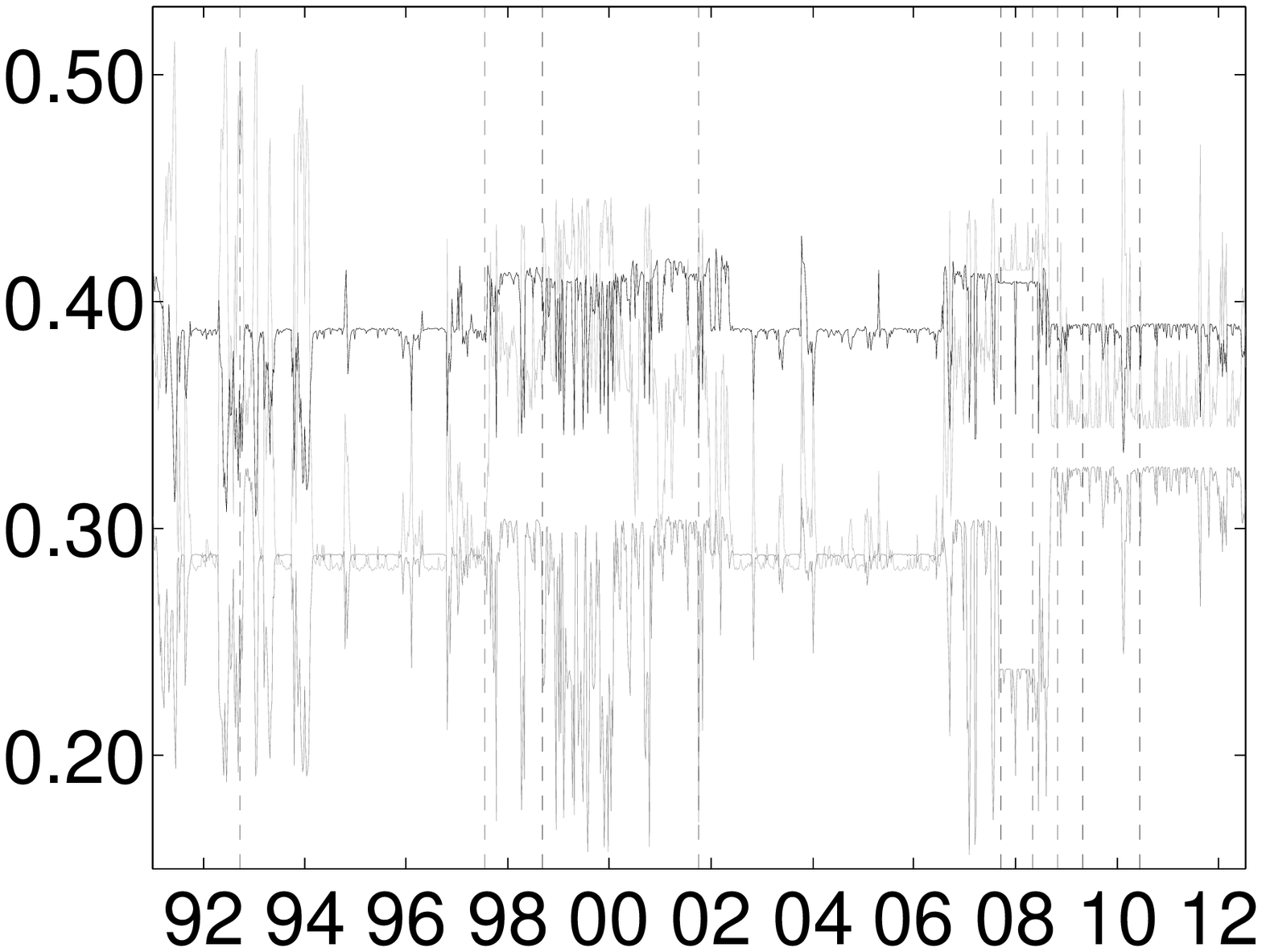}
\caption{\footnotesize{Banks \textit{(light gray)}, financial services \textit{(gray)} and non--life insurance \textit{(dark)} against life insurance.}}
\label{fig:SV_CoES_life}
\end{subfigure}\qquad\qquad
\begin{subfigure}[b]{0.3\textwidth}
\includegraphics[width=1.0\linewidth, height=0.6\linewidth]{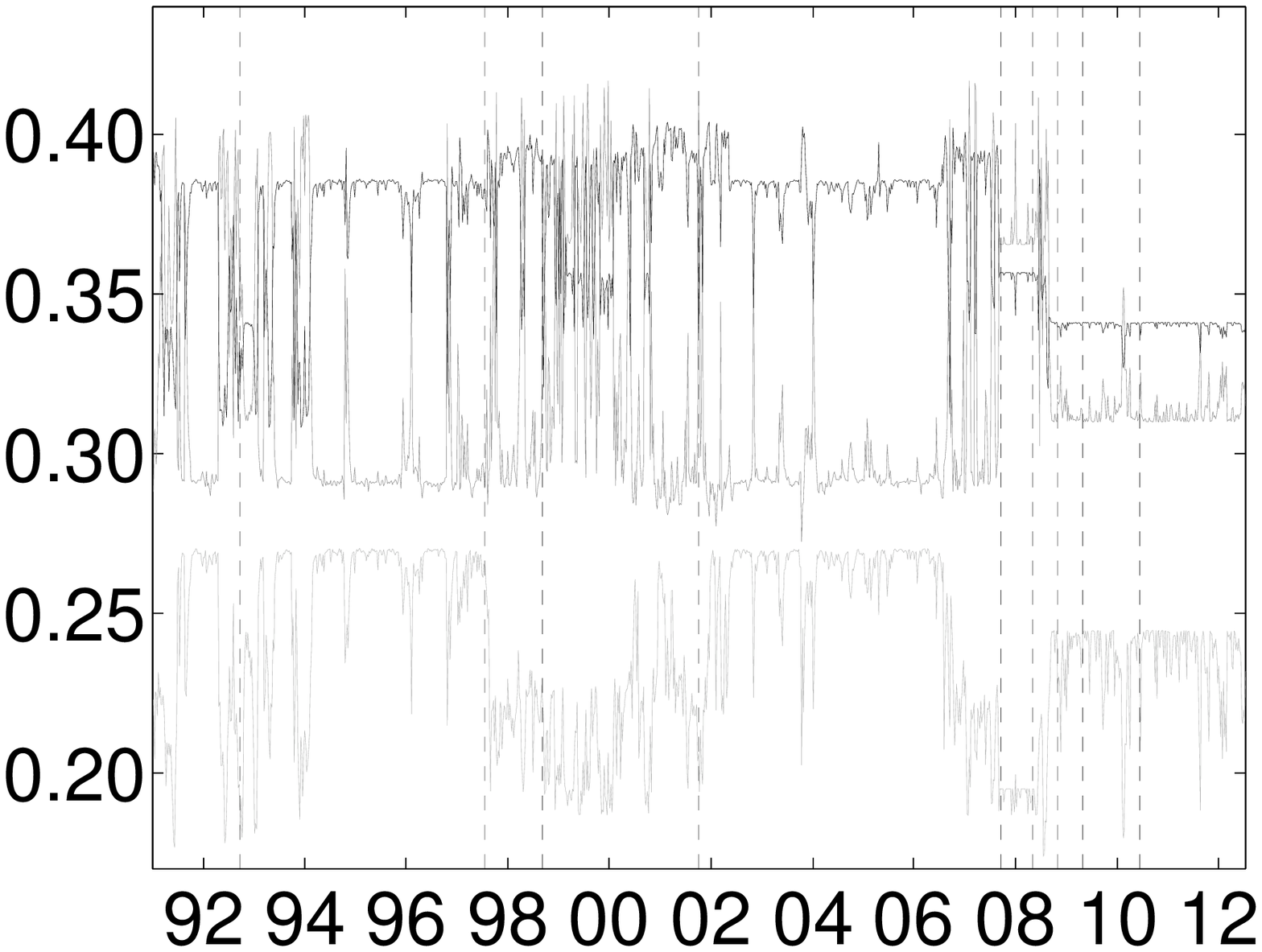}
\caption{\footnotesize{Banks \textit{(light gray)}, financial services \textit{(gray)} and life insurance \textit{(dark)} against non--life insurance.}}
\label{fig:SV_CoES_non-life}
\end{subfigure}
\captionsetup{font={footnotesize}, labelfont=sc}
\caption{\footnotesize{Shalpey value based on $\Delta^\sM$--CoES of the different sectors against banks (\subref{fig:SV_CoES_banks}), financial services (\subref{fig:SV_CoES_financials}), life insurance (\subref{fig:SV_CoES_life}) and non--life insurance (\subref{fig:SV_CoES_non-life}).
Vertical dotted lines represent major financial downturns: the ``Black Wednesday'' (September 16, 1992), the Asian crisis (July, 1997), the Russian crisis (August, 1998), the September 11, 2001 shock, the Bear Stearns hedge funds collapse (August 5, 2007), the Bear Stearns acquisition by JP Morgan Chase, (March 16, 2008), the Lehman's failure (September 15, 2008), the peak of the onset of the recent global financial crisis (March 9, 2009) and the European sovereign-debt crisis of April 2010 (April 23, 2010, Greek crisis).}}
\label{fig:USSectors_Shapley_value_CoES_ALL}
\end{center}
\end{figure}
%

%
\begin{figure}[!t]
\begin{center}
\begin{subfigure}[b]{0.3\textwidth}
\includegraphics[width=1.0\linewidth, height=0.6\linewidth]{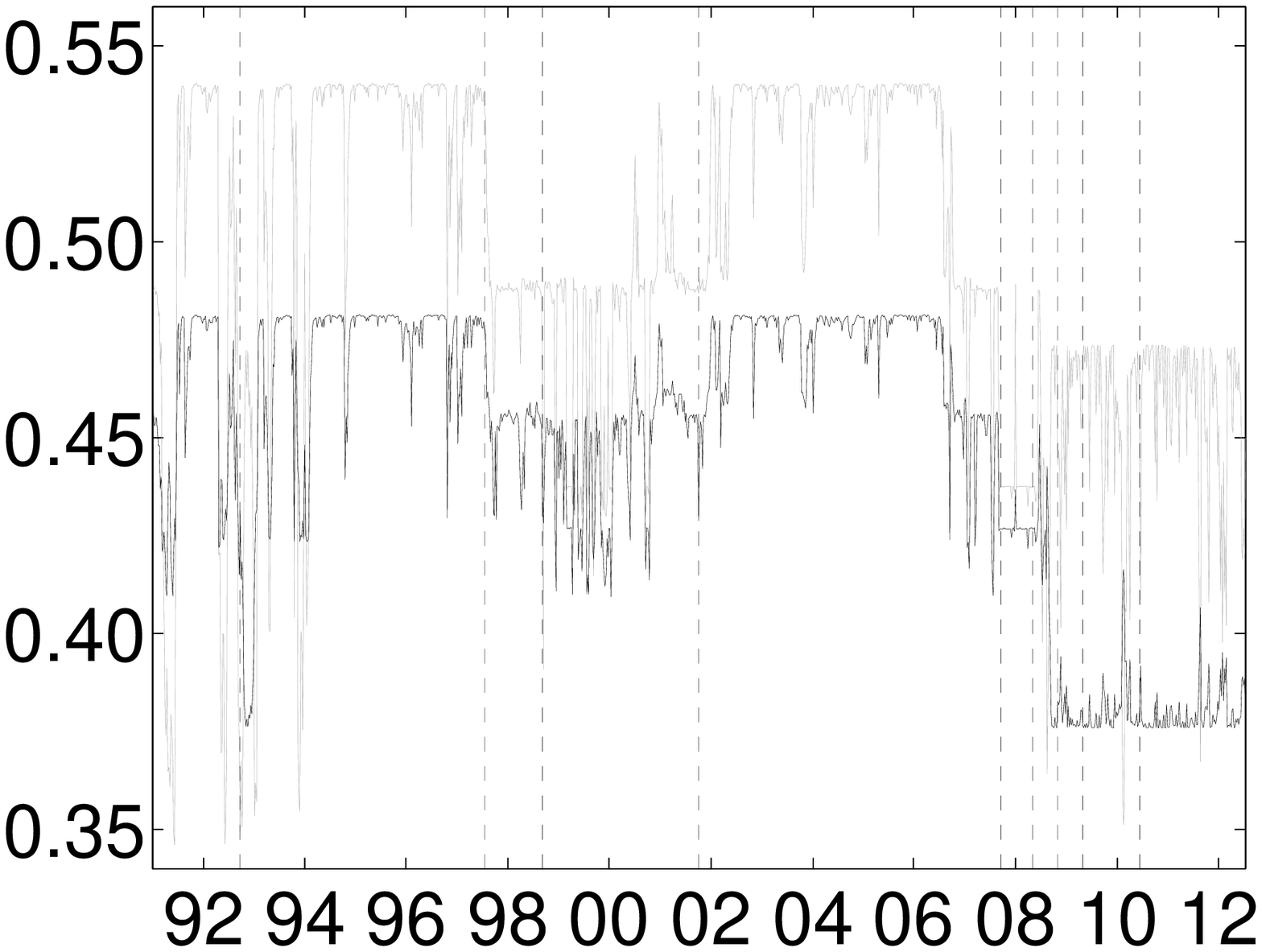}
\caption{\footnotesize{Distress of banks on financial services \textit{(dark)} and viceversa \textit{(light gray)}.}}
\label{fig:SV_CoVaR_banks_vs_fin_srvs}
\end{subfigure}\qquad\qquad
\begin{subfigure}[b]{0.3\textwidth}
\includegraphics[width=1.0\linewidth, height=0.6\linewidth]{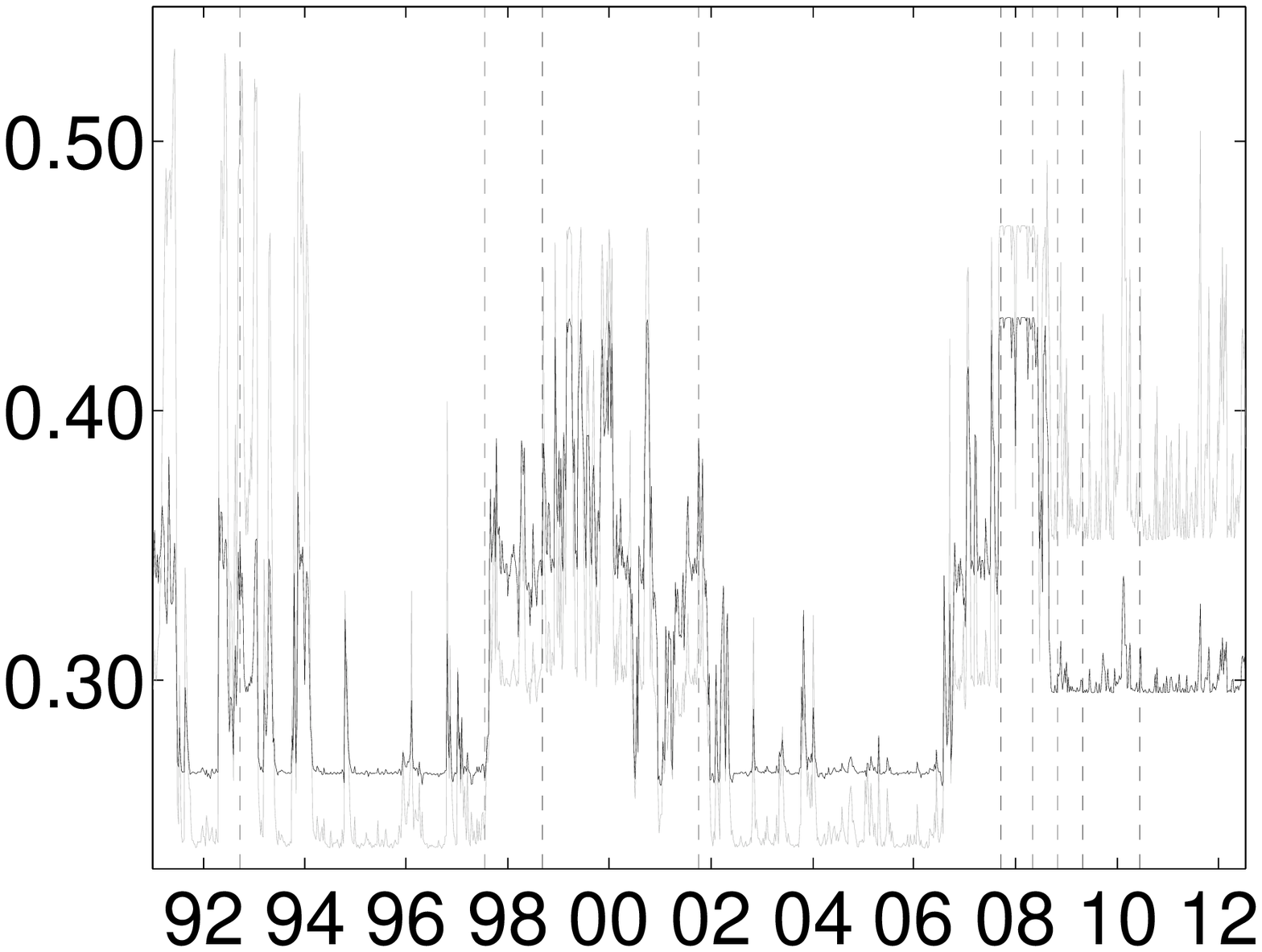}
\caption{\footnotesize{Distress of banks on life insurance \textit{(dark)} and viceversa \textit{(light gray)}.}}
\label{fig:SV_CoVaR_banks_vs_life}
\end{subfigure}\\
\vspace{1.0cm}
\begin{subfigure}[b]{0.3\textwidth}
\includegraphics[width=1.0\linewidth, height=0.6\linewidth]{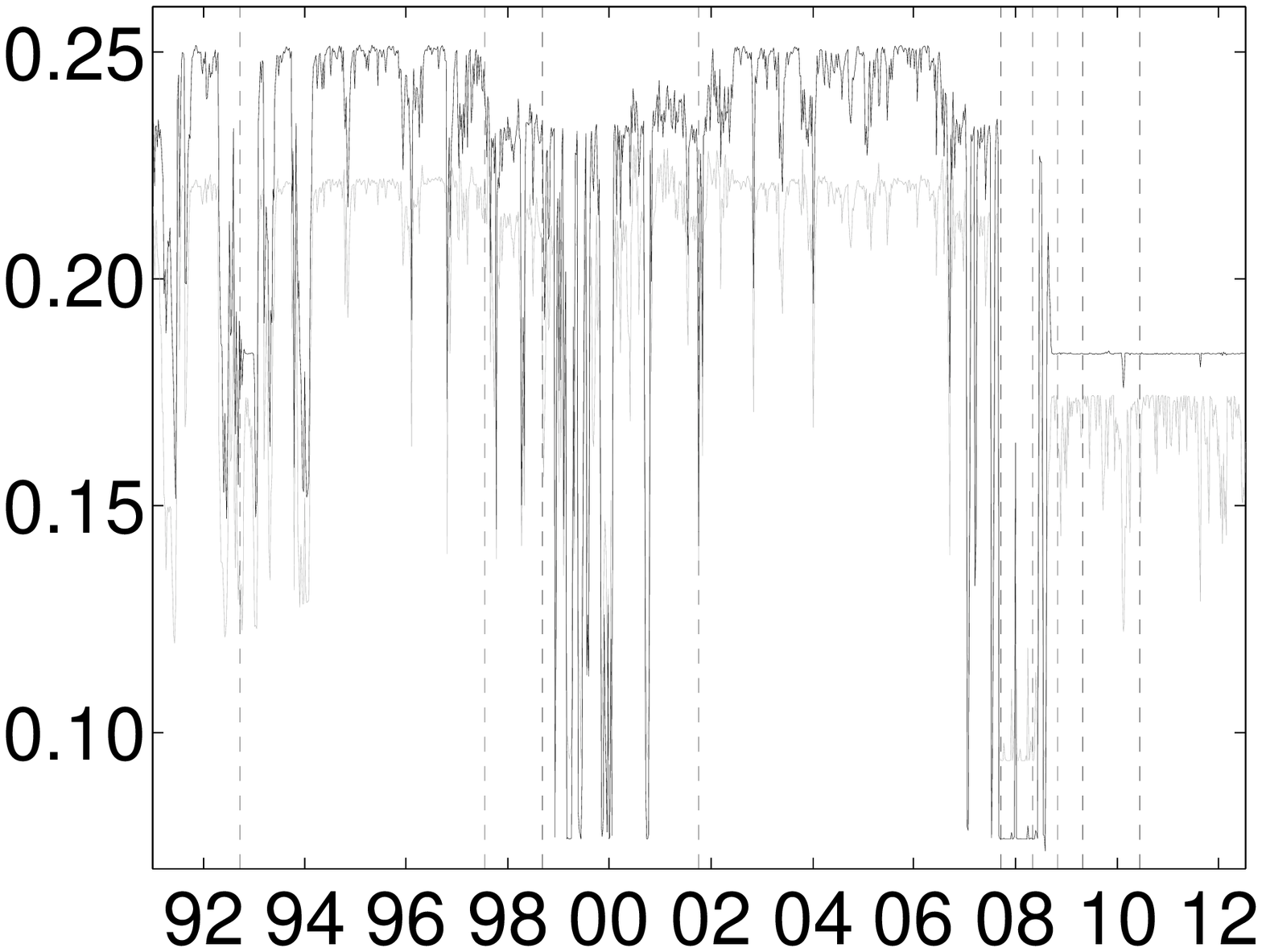}
\caption{\footnotesize{Distress of banks on non--life insurance \textit{(dark)} and viceversa \textit{(light gray)}.}}
\label{fig:SV_CoVaR_banks_vs_non-life}
\end{subfigure}\qquad\qquad
\begin{subfigure}[b]{0.3\textwidth}
\includegraphics[width=1.0\linewidth, height=0.6\linewidth]{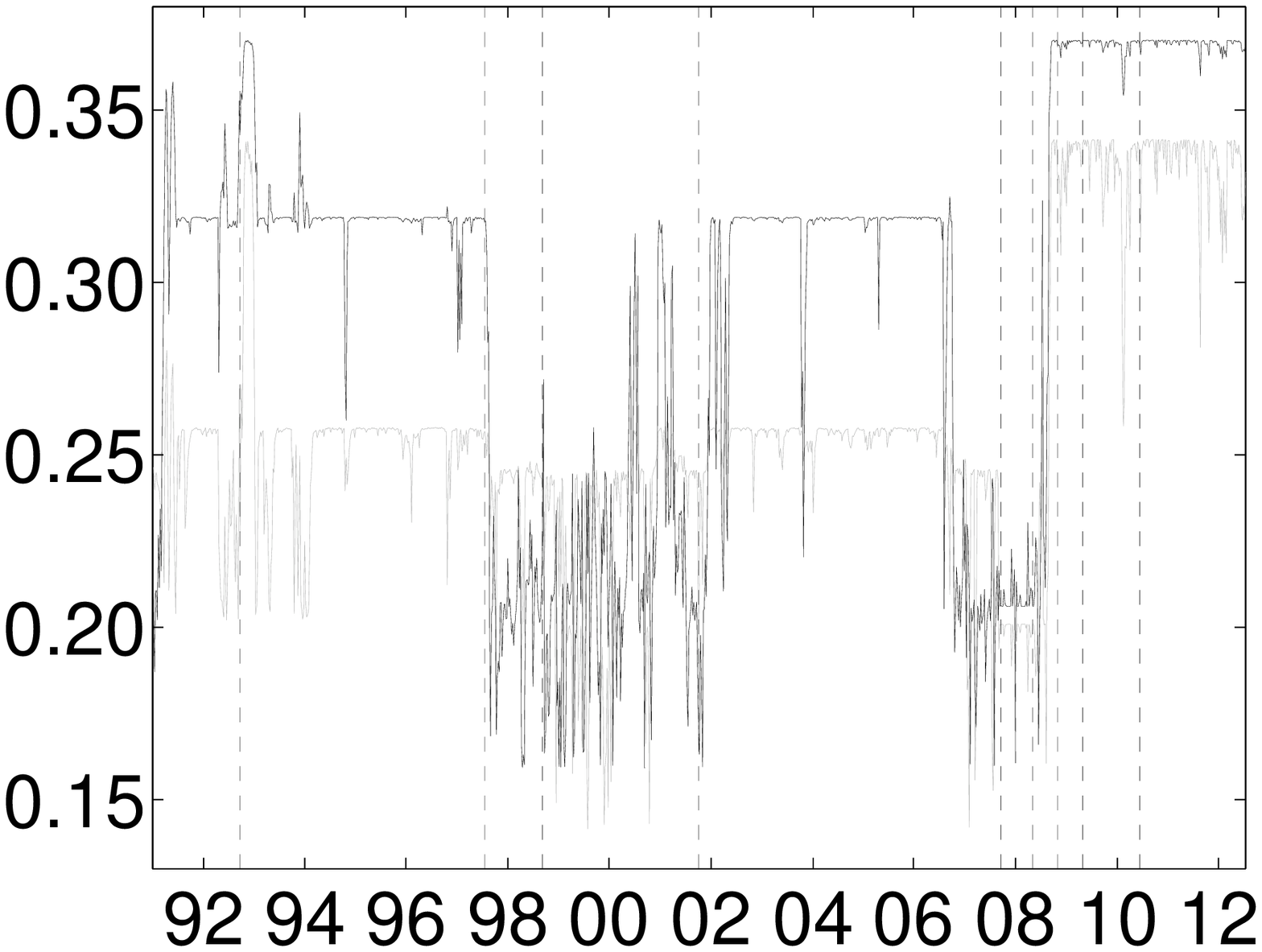}
\caption{\footnotesize{Distress of financial srvs on life insurance \textit{(dark)} and viceversa \textit{(light gray)}.}}
\label{fig:SV_CoVaR_fin_srvs_vs_life}
\end{subfigure}\\
\vspace{1.0cm}
\begin{subfigure}[b]{0.3\textwidth}
\includegraphics[width=1.0\linewidth, height=0.6\linewidth]{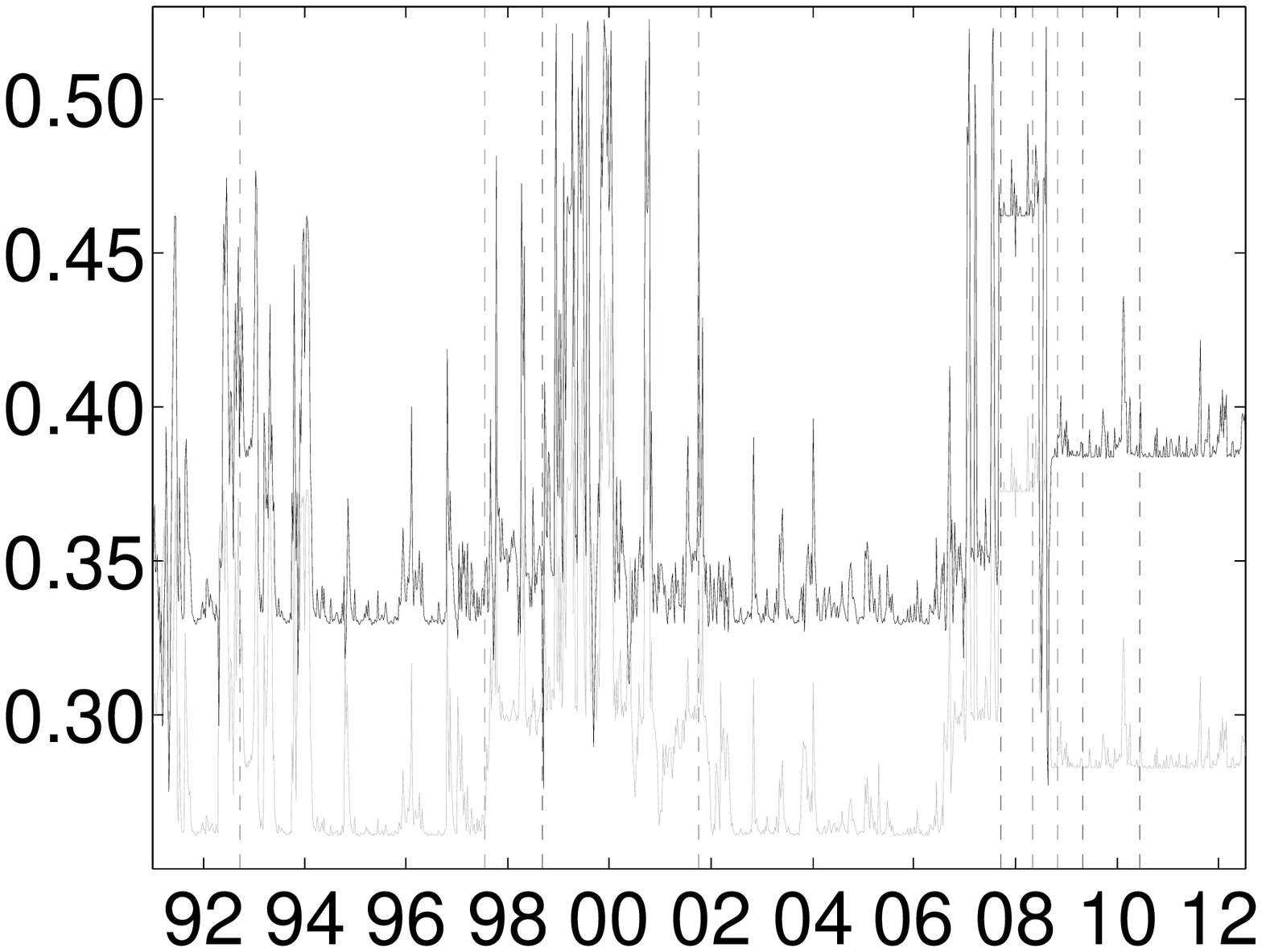}
\caption{\footnotesize{Distress of financial srvs on non--life insurance \textit{(dark)} and viceversa \textit{(light gray)}.}}
\label{fig:SV_CoVaR_fin_srvs_vs_non-life}
\end{subfigure}\qquad\qquad
\begin{subfigure}[b]{0.3\textwidth}
\includegraphics[width=1.0\linewidth, height=0.6\linewidth]{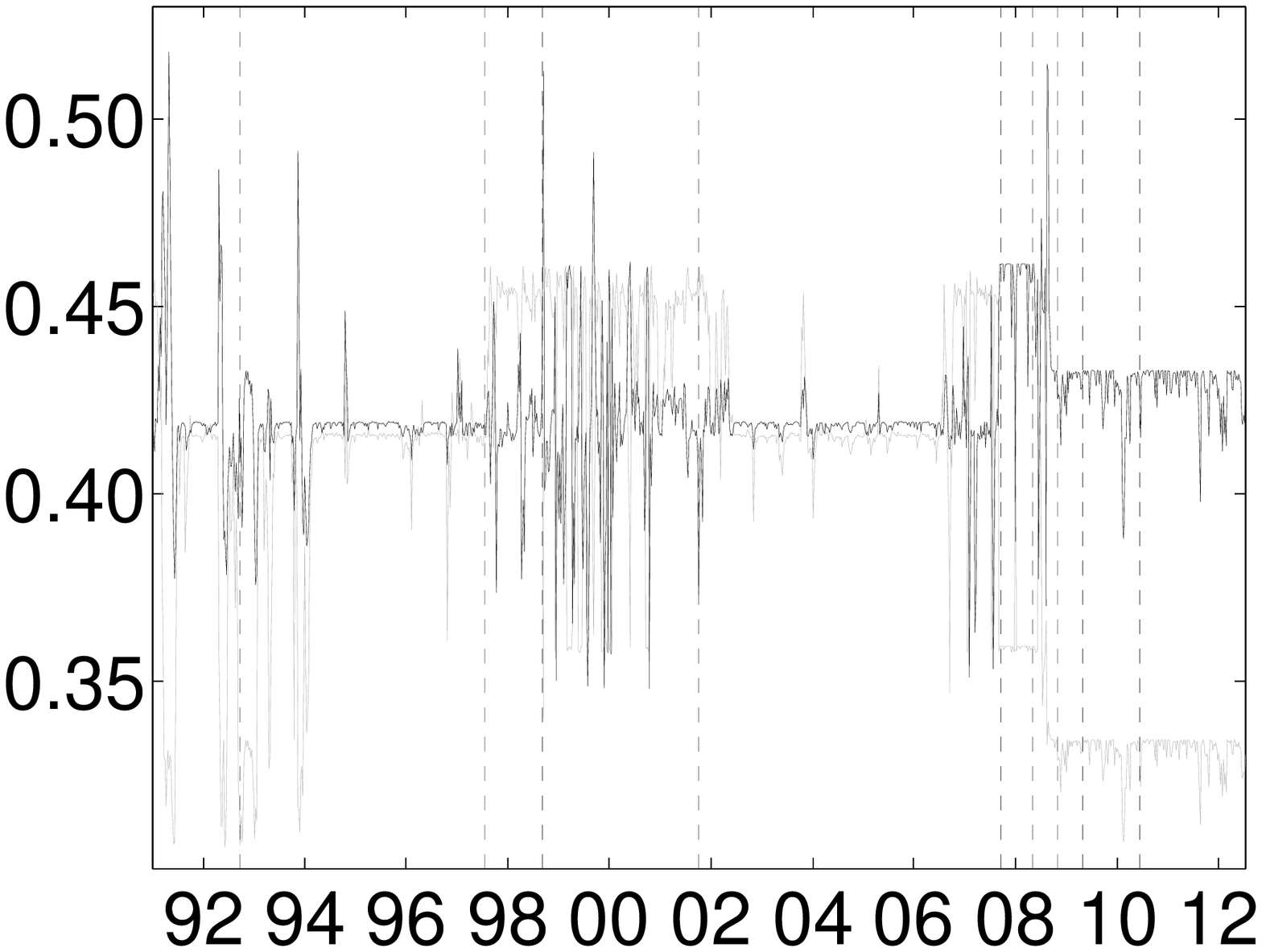}
\caption{\footnotesize{Distress of life insurance on non--life insurance \textit{(dark)} and viceversa \textit{(light gray)}.}}
\label{fig:SV_CoVaR_life_vs_non-life}
\end{subfigure}
\captionsetup{font={footnotesize}, labelfont=sc}
\caption{\footnotesize{Vis--\`a--vis comparisons of Shapleys value based on $\Delta^\sM$CoVaR for the different sectors.
Vertical dotted lines represent major financial downturns: the ``Black Wednesday'' (September 16, 1992), the Asian crisis (July, 1997), the Russian crisis (August, 1998), the September 11, 2001 shock, the Bear Stearns hedge funds collapse (August 5, 2007), the Bear Stearns acquisition by JP Morgan Chase, (March 16, 2008), the Lehman's failure (September 15, 2008), the peak of the onset of the recent global financial crisis (March 9, 2009) and the European sovereign-debt crisis of April 2010 (April 23, 2010, Greek crisis).}}
\label{fig:USSectors_Shapley_value_CoVaR_2by2}
\end{center}
\end{figure}
%

%
\begin{figure}[!t]
\begin{center}
\begin{subfigure}[b]{0.3\textwidth}
\includegraphics[width=1.0\linewidth, height=0.6\linewidth]{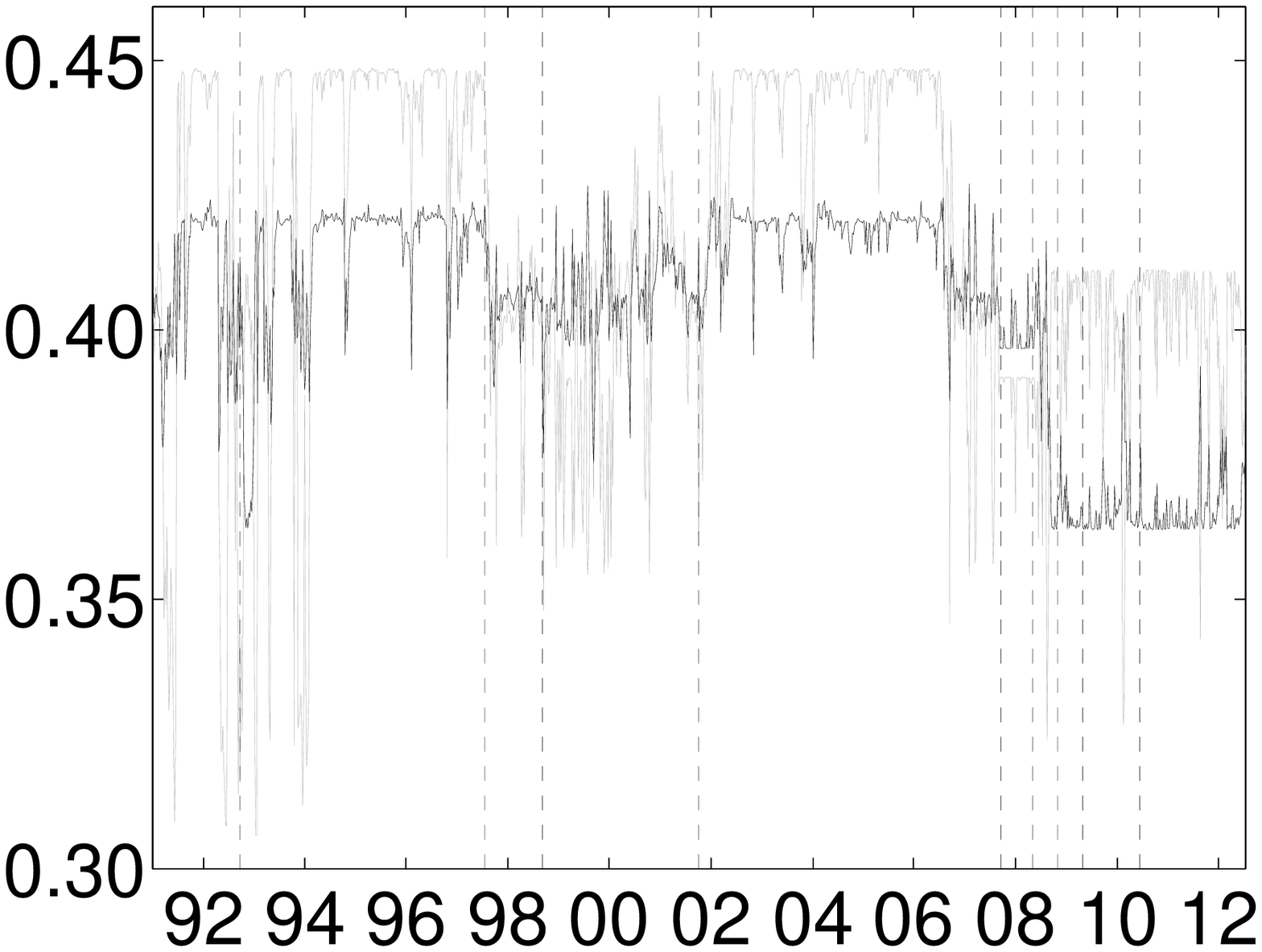}
\caption{\footnotesize{Distress of banks on financial services \textit{(dark)} and viceversa \textit{(light gray)}.}}
\label{fig:SV_CoES_banks_vs_fin_srvs}
\end{subfigure}\qquad\qquad
\begin{subfigure}[b]{0.3\textwidth}
\includegraphics[width=1.0\linewidth, height=0.6\linewidth]{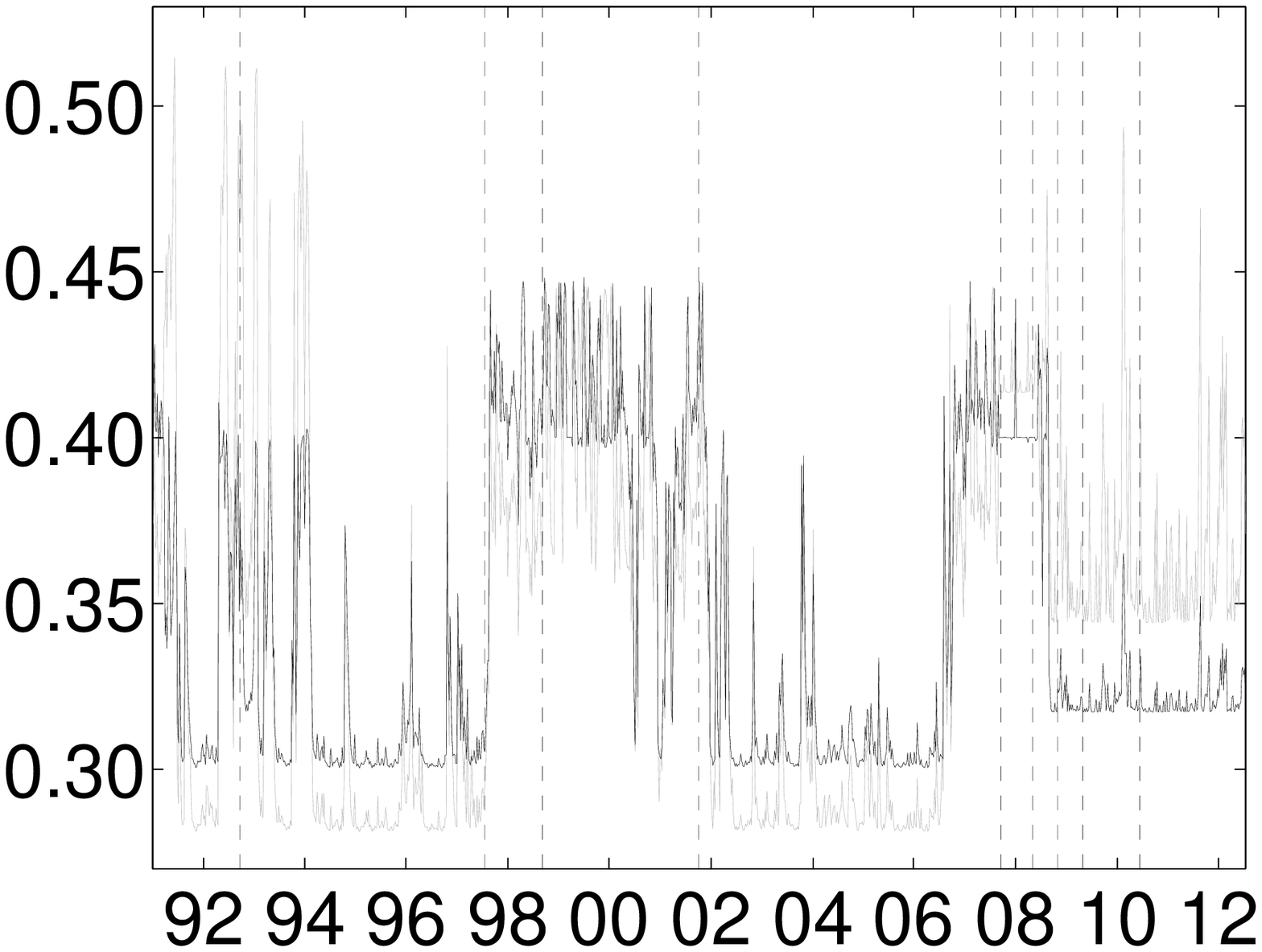}
\caption{\footnotesize{Distress of banks on life insurance \textit{(dark)} and viceversa \textit{(light gray)}.}}
\label{fig:SV_CoES_banks_vs_life}
\end{subfigure}\\
\vspace{1.0cm}
\begin{subfigure}[b]{0.3\textwidth}
\includegraphics[width=1.0\linewidth, height=0.6\linewidth]{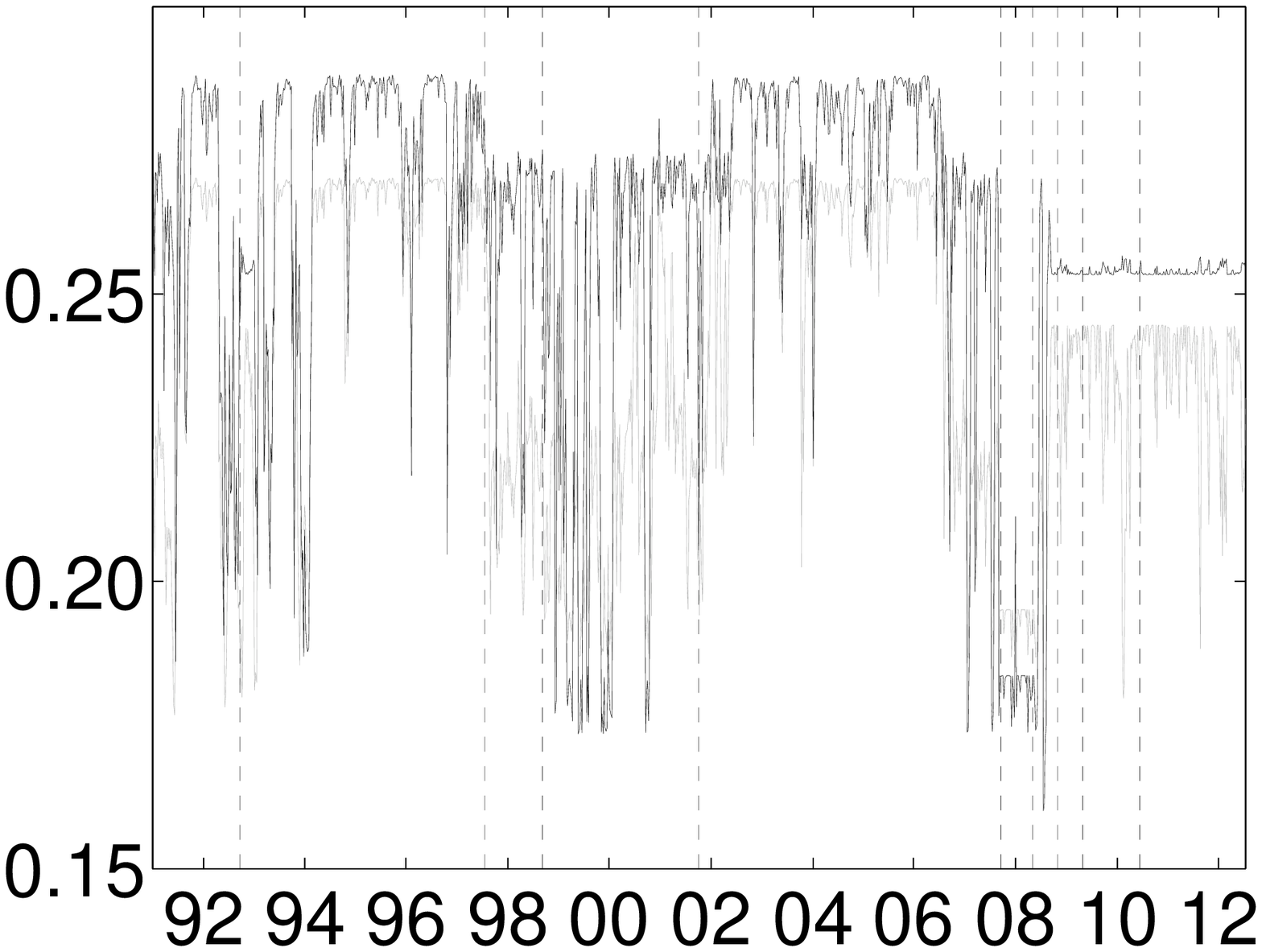}
\caption{\footnotesize{Distress of banks on non--life insurance \textit{(dark)} and viceversa \textit{(light gray)}.}}
\label{fig:SV_CoES_banks_vs_non-life}
\end{subfigure}\qquad\qquad
\begin{subfigure}[b]{0.3\textwidth}
\includegraphics[width=1.0\linewidth, height=0.6\linewidth]{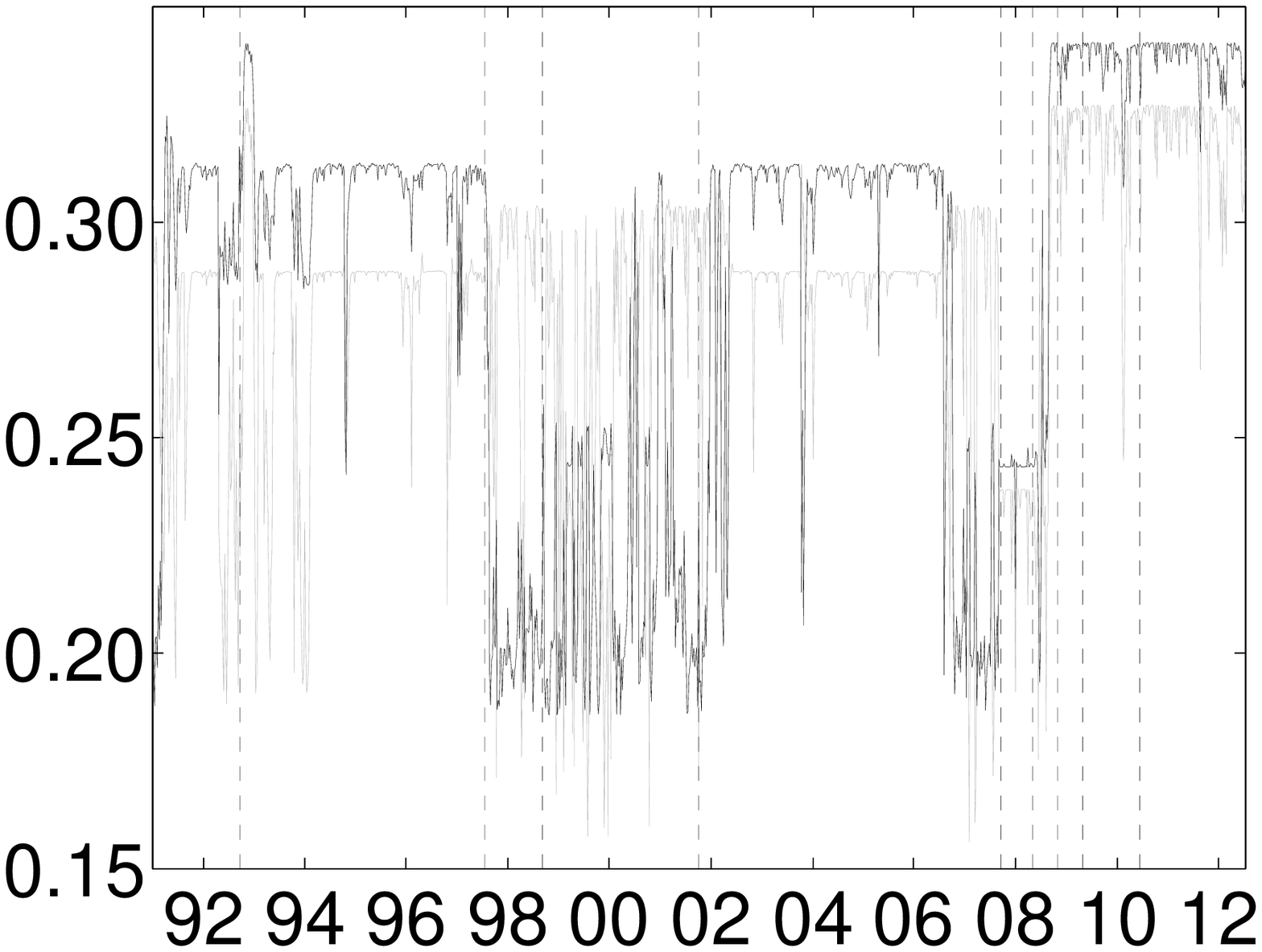}
\caption{\footnotesize{Distress of financial srvs on life insurance \textit{(dark)} and viceversa \textit{(light gray)}.}}
\label{fig:SV_CoES_fin_srvs_vs_life}
\end{subfigure}\\
\vspace{1.0cm}
\begin{subfigure}[b]{0.3\textwidth}
\includegraphics[width=1.0\linewidth, height=0.6\linewidth]{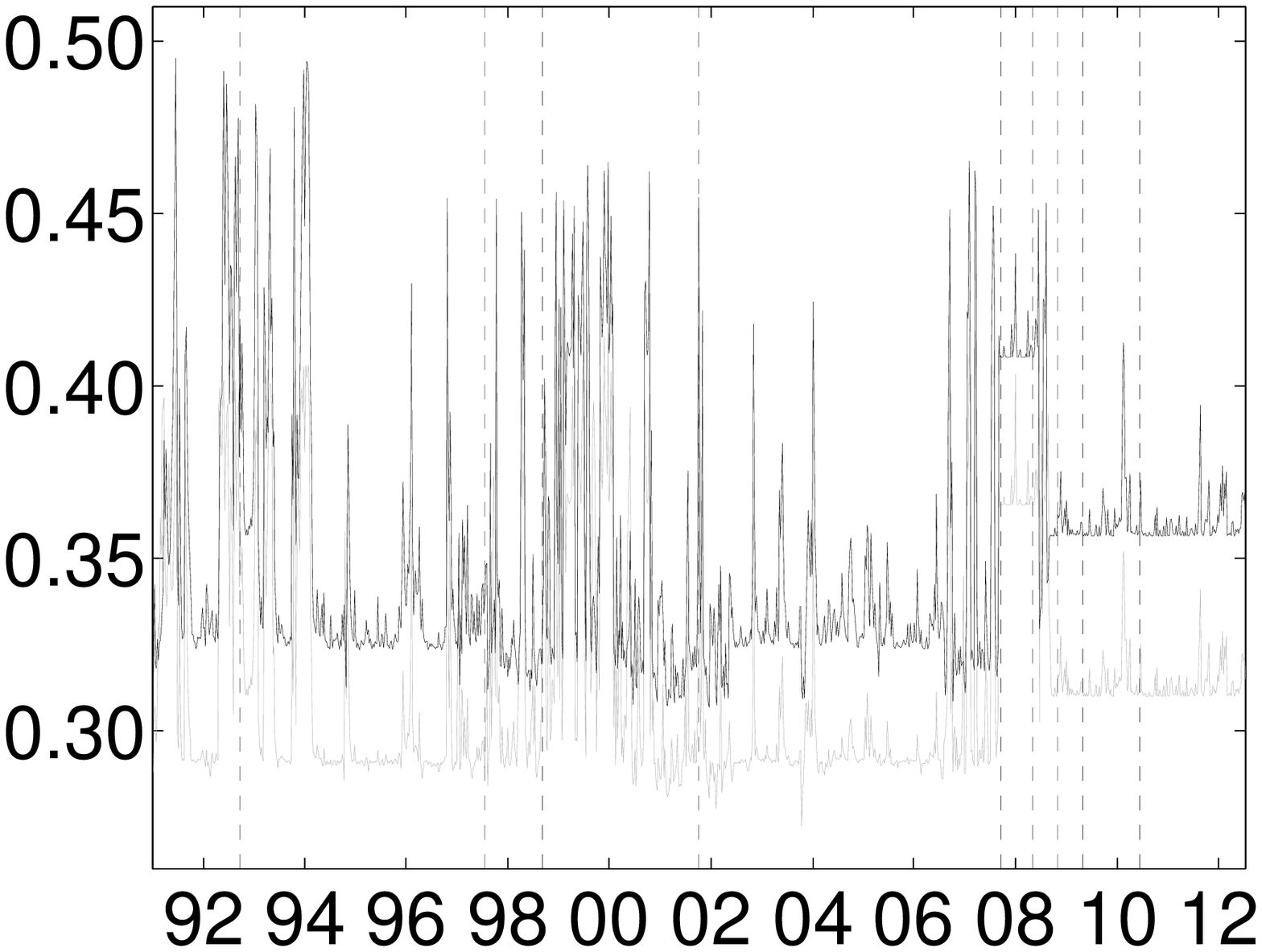}
\caption{\footnotesize{Distress of financial srvs on non--life insurance \textit{(light gray)} and viceversa \textit{(dark)}.}}
\label{fig:SV_CoES_fin_srvs_vs_non-life}
\end{subfigure}\qquad\qquad
\begin{subfigure}[b]{0.3\textwidth}
\includegraphics[width=1.0\linewidth, height=0.6\linewidth]{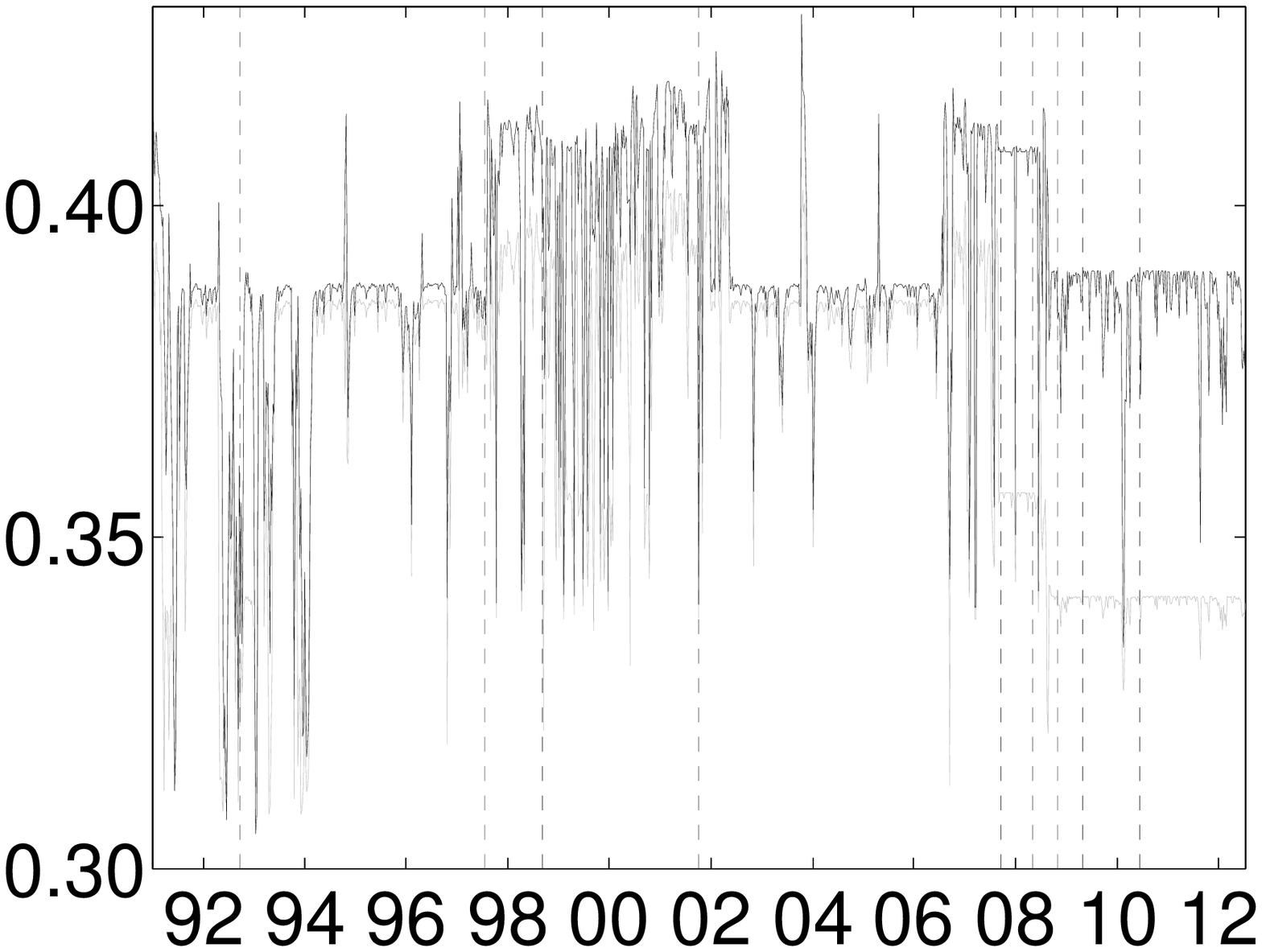}
\caption{\footnotesize{Distress of life insurance on non--life insurance \textit{(light gray)} and viceversa \textit{(dark)}.}}
\label{fig:SV_CoES_life_vs_non-life}
\end{subfigure}
\captionsetup{font={footnotesize}, labelfont=sc}
\caption{\footnotesize{Vis--\`a--vis comparisons of Shapleys value based on $\Delta^\sM$CoES for the different sectors.
Vertical dotted lines represent major financial downturns: the ``Black Wednesday'' (September 16, 1992), the Asian crisis (July, 1997), the Russian crisis (August, 1998), the September 11, 2001 shock, the Bear Stearns hedge funds collapse (August 5, 2007), the Bear Stearns acquisition by JP Morgan Chase, (March 16, 2008), the Lehman's failure (September 15, 2008), the peak of the onset of the recent global financial crisis (March 9, 2009) and the European sovereign-debt crisis of April 2010 (April 23, 2010, Greek crisis).}}
\label{fig:USSectors_Shapley_value_CoES_2by2}
\end{center}
\end{figure}
%
\indent The goal of this section is to analyse how the overall risk spreads among different sectors by inspecting the time evolution of the Shapley values risk measures introduced in the previous sections. For further details on the application of the Shapley value as a measure of risk attribution and its properties we refer the reader to the paper of Bernardi \textit{et al.} \citeyearpar{bernardi_etal.2013b}.\newline
\indent We examine whether the extreme--tail interdependence among banks, financial services and insurance sectors have changed over time. The interconnectedness among sectors, and in particular between the financial and the insurance sectors, have been highly investigated in the recent literature. In their empirical investigation, Bernal \textit{et al.} \citeyearpar{bernal_etal.2013}, for example, found that banks contribute relatively the most to systemic risk in the Eurozone, while the insurance industry is the most systemically risky sector in the US for the period 2004--2012. Recently, Chen \textit{et al.} (2013) and Billio \textit{et al.} \citeyearpar{billio_etal.2012}, using univariate Granger--causality analyses, show a significant two--way interconnection between banking and insurance sectors. Our model--based approach instead is able to investigate the contribution of each sector to the risk of all the remaining ones in a multivariate framework. Furthermore our analysis considers different distress events jointly affecting the market participants health level. 
\indent Figure \ref{fig:USSectors_Total_Risk} plots the overall risk of each sector based on Multiple--CoVaR and Multiple--CoES at confidence level $\tau_1=\tau_2=0.05$. The total risk in Figure \ref{fig:USSectors_Total_Risk} provides the benchmark for regulators to calculate individual sector contributions to the overall risk when all the remaining others are in distress. The dynamic evolution of the total risk for all sectors suggests that, during the analysed period, there have been four major downside peaks: the Russian crisis at the end 1998 which partially overlaps the dot--com bubble of 1999--2000; the 11--September shock and the recent global financial crisis of 2007--2009. Interestingly, both the Russian crisis and the recent mortgage subprime crisis of 2007, that are followed by several years of financial turbulence affecting all the sectors, are anticipated by a long period where the total risk increased significantly. Although both crisis episodes have been characterised different durations as well as financial and economic conditions we observe that they are anticipated by an increased level of overall risk. This evidence supports the use of the total risk measure as a leading indicator for the financial crisis. Concerning in particular the Russian crisis, a careful inspection of Figure \ref{fig:USSectors_Total_Risk} we observe that the total risk for all sectors increased suddenly in conjunction with the antecedent Asian crisis of middle 1997. Then, the system experienced a long period of financial instability encompassing all the three subsequent crisis culminated with the Twin towers attack of September 2001. Concerning the recent global crisis, instead, we observe that the total risk increases long time before the Bear Stearns hedge found collapse of August 2007, for all sectors. In fact, looking at Figure \ref{fig:USSectors_SmoothProb} we note that the system transits into state 2 since mid 2006 anticipating the subsequent market turbulence.
Another important difference between the two periods of financial crisis, 1997--2001 and 2007--2009, emerges by comparing the total risk evolution in Figure \ref{fig:USSectors_Total_Risk} with the smoothed probabilities in Figure \ref{fig:USSectors_SmoothProb}. It is evident that the latter financial crisis has been more persistent than those occurred in the previous decades. In fact, during the period 1997--2001 the system is in state 2 of moderate instability while during the period 2007--2009 the system is in state 1 of financial turbulence for most of the time. Finally, we observed that the US market is not affected by European sovereign debt crisis in May 2010.\newline
\indent It is interesting to note that the total risk dynamics in Figure \ref{fig:USSectors_Total_Risk} suggests that, before the 2007, the financial services sector (red line) is the most affected by the other sectors' distress, while the banking and the insurance sectors display a similar low level of risk, whereupon this ordering completely change by the beginning of the 2009, with banks (blue line) and life insurance (dark line) being the most risky sectors. In fact, we observe that the total risk contributions change during crisis periods becoming larger in levels and reversing the ordering of importance.\newline
\indent As said before the total risk contribution analysis identifies the sector most affected by the crisis of all the remaining ones during different periods of time providing an important tool for risk management. This analysis by itself is not conclusive because it considers all the other institutions or sectors except one at their distress level. This is the reason why, once we get the $\Delta^\sM$CoVaR and $\Delta^\sM$CoES risk measures for each sector given all the possible combinations of the remaining sectors distress' events, we apply the Shapley value methodology to compose the puzzle of their synthesis to provide a unifying measure. The resulting Shapley values act as an overall risk distributor among the market participants providing the marginal contributions to each of the considered sectors of the remaining sectors' distress. In our case, since we are measuring interconnections among four sectors, we have three Shapley value contributions for each of them. Figure \ref{fig:USSectors_Shapley_value_CoVaR_ALL} and \ref{fig:USSectors_Shapley_value_CoES_ALL} plot the individual marginal contributions calculated by means of the Shapley value ${\rm ShV}_i$ based on $\Delta^{\sM}{\rm CoVaR}_{i\vert\mathcal{J}_\sd}$ and $\Delta^{\sM}{\rm CoES}_{i\vert\mathcal{J}_\sd}$, respectively. Figure (\ref{fig:SV_CoVaR_banks}) plots the Shapley Values contributions of financial services, life insurance and non--life insurance on the banking sector. We observe that the financial services (light gray line) sector contributes more to the distress of the banking sector than both the life and non--life insurance sectors, independently on the overall financial situation identified by the hidden state. Furthermore, during  periods of financial stability, the contribution of financial services is about two times larger than that of life and non--life insurance. During periods of crisis, instead, the contributions of financial services and non--life insurance decrease, while that of life insurance increases, so that the contribution of non--life insurance becomes negligible and those of financial services and life insurance approach almost the same level. Figure (\ref{fig:SV_CoVaR_financials}) plots the contributions on the financial services sector. The picture is similar to the one in Figure (\ref{fig:SV_CoVaR_banks}) with the role of banks and financial services and that of life and non--life insurance reversed. This essentially means that banks and financial services are strongly interconnected. Figure (\ref{fig:SV_CoVaR_life}) and (\ref{fig:SV_CoVaR_non-life}) plot the contributions on life and non--life insurance sectors, respectively. Comparing the two figures we observe that these sectors are highly interconnected during period of crisis as well as during period of financial stability. In Figure (\ref{fig:SV_CoVaR_life}) the contributions of banks and financial services on life insurance looks the same and they increase suddenly after the recent global financial crisis ends. In Figure (\ref{fig:SV_CoVaR_non-life}) the contribution of banks on non--life insurance is low compared to other sectors' contributions, suggesting that the banking sector is more interconnected with the life insurance (see Figure \ref{fig:SV_CoVaR_life}) than with the non--life one. The results here presented are confirmed by inspecting Figure \ref{fig:USSectors_Shapley_value_CoES_ALL} where individual marginal contributions are calculated by means of the Shapley value ${\rm ShV}_i$ based on $\Delta^{\sM}{\rm CoES}_{i\vert\mathcal{J}_\sd}$.\newline
\indent To deeply understand the relative impact of each sector's financial distress on all the other sectors, we plot the two way comparisons of marginal contributions for all the possible pairs of sectors. Figures \ref{fig:USSectors_Shapley_value_CoVaR_2by2} and \ref{fig:USSectors_Shapley_value_CoES_2by2} plot the vis--\`a--vis comparisons based on the Shapley value $\Delta^\sM$CoVaR and $\Delta^\sM$CoES, respectively. Figures \ref{fig:SV_CoVaR_banks_vs_fin_srvs} and \ref{fig:SV_CoES_banks_vs_fin_srvs} consider the banking and diversified financial services sectors. Independently of the hidden state, it is clear that the financial services' distress impacts more the banking sector than the other way around.
Figures \ref{fig:SV_CoVaR_banks_vs_life} and \ref{fig:SV_CoES_banks_vs_life} consider the banking sector against the life insurance one. This picture highlights an important aspect of the extreme interconnection between these sectors that can be captured by the dynamic Shapley approach here proposed. In particular, it is evident that, prior to the recent global financial crisis of 2007--2009, banks impact more the life insurance during periods of financial stability (for example 2002--2006), than the vice--versa, while, after the end of 2008, we observe the order of importance between these two sectors reverses. This behaviour is peculiar of the banking and life insurance sectors, and does not characterise the relationship between the banking sector on non--life insurance depicted in Figures \ref{fig:SV_CoVaR_banks_vs_non-life} and \ref{fig:SV_CoES_banks_vs_non-life}. These two latter sectors seem to be highly interconnected with a slight predominance of the banks.   
Figures \ref{fig:SV_CoVaR_banks_vs_non-life} and \ref{fig:SV_CoES_banks_vs_non-life}  consider the impact of the financial sector on life and non--life insurance sectors, providing clear evidence that the financial sector highly impacts the insurance sector. Moreover, the $\Delta^\sM$CoVaR and $\Delta^\sM$CoES Shapley values seem to be slightly different. In fact, if we look at the $\Delta^\sM$CoVaR--Shapley value in Figure \ref{fig:SV_CoVaR_fin_srvs_vs_life} it seems that, during periods of financial instability, life insurance provides a larger contribution than financial services. Finally, Figures \ref{fig:SV_CoVaR_life_vs_non-life} and \ref{fig:SV_CoES_life_vs_non-life} consider the interconnection between the insurance sectors, providing clear evidence that these two sectors are highly interconnected till the end of 2008. After the end of 2008 relative weight of life insurance decreased.\newline
\indent In Table \ref{fig:USSector_Shapley_value_CoES}, we provide the comparison between the estimated Shapley--$\Delta^\sM$CoES series and the Standard $\Delta$CoES series for each of the considered sectors against each other. For each pair of sectors, the standard $\Delta$CoES of Adrian and Brunnermeier \citeyearpar{adrian_brunnermeier.2011} is calculated appropriately modifying equation \eqref{eq:delta_m_es} on the bivariate Student--t MS output. Is is evident that, as expected, except for the banks and life insurance pairs, the signal provided by the standard $\Delta$CoES differs significantly from that provided by our risk measure based on the Shapley value. Ad discussed by Bernardi \textit{et al.} \citeyearpar{bernardi_etal.2013b} the two risk measures coincide only under the conditional independence assumption for the Student--t case.
%
\section{Discussion}
\label{sec:discussion}
%
The model--based approach to the overall risk assessment developed in Bernardi \textit{et al.} \citeyearpar{bernardi_etal.2013b} and applied here to four sector belonging to the US Dow Jones Composite Index allows to understand how the risk spread among the banking, life and non--life insurance and other financial services sectors. Our empirical findings suggest that each financial sector significantly impacts each other during crisis periods, as well as during more stable phases. When comparing the contribution of each financial industry, banks appear to be the major source of risk for all the remaining ones, followed by the financial services and the insurance sectors. This results are in line with previous findings (see e.g. Bernal \textit{et al.}, \citeyear{bernal_etal.2013}) and are supported by theoretical arguments. Adams \textit{et al.} \citeyearpar{adams_etal.2011} and Girardi and Erg\"un \citeyearpar{girardi_ergun.2013}, for example, found that the banking sector is systemically riskier than the insurance sector. Banks are particularly fragile institutions because of their core business -- especially credit activity to households and corporate companies along with short--term funding -- making the banking sector particularly exposed to the overall risk. insurance companies, instead, generally fund themselves through long--term premiums and have higher disposable--liquidity. This latter argument plays an important role since financial crises usually begin with a liquidity squeeze that further weakens the capital position of vulnerable firms. Another argument is that balance sheets of banks are highly volatile and exposed to economic cycles while insurance companies usually present simple and economically stable balance sheets due to their long--term oriented business.\newline
\indent Another aspect of relevant interest is the degree of financial interconnection among different financial sectors with particular emphasis to the banking and insurance ones. Our Shapley value $\Delta^\sM$CoVaR and $\Delta^\sM$CoES risk measures are able to identify sector interconnections as well as the directions through which the spillover mechanism takes place. Recently, the financial literature has pointed out the fact that banks are also much more interconnected than insurance companies through interbank lending such as the repo market. Eventually, the size of the banking sector much greater than the one of the insurance sector can also be a factor explaining that banks appear as the most risky in our study.\newline 
\indent Finally, our results regarding the systemic role of the four financial sectors in the United States are consistent with recent argument raised in the literature emphasising the risk associated to fast growing non--core activities (such as credit derivatives) of insurance companies (Bell and Keller, \citeyear{bell_keller.2009}; Cummins and Weiss, \citeyear{cummins_weiss.2012}). In fact, he non-core activities of U.S. insurance firms highly increased over the last decade. 
%
%
\section{Conclusion}
\label{sec:conclusion}
%
This paper aims to assess the contribution of the different sectors of the financial system to the overall risk and to measure their degree of interconnectedness. To that end, we split the financial system into four sectors corresponding respectively to the banking, life and non--life insurance and other financial services industries. The impact of distress within any one of these sectors is measured using the Shapley value $\Delta^\sM$CoVaR (or $\Delta^\sM$CoES) risk measure proposed by Bernardi \textit{et al.} \citeyearpar{bernardi_etal.2013b}. More precisely, the Shapley value $\Delta^\sM$CoVaR ($\Delta^\sM$CoES) extends the traditional $\Delta$CoVaR approach of Adrian and Brunnermeier \citeyearpar{adrian_brunnermeier.2011} to the case where multiple distress events are jointly observed as it is often the case during period of financial instability. The Shapley value $\Delta^\sM$CoVaR and $\Delta^\sM$CoES can therefore be interpreted as the additional level of risk faced by each sector arising from the distress of one or more of the remaining financial sectors of interest. Empirical results reveal that in the US financial market, for the period ranging from 1992 to 2012, the banking sector contributes relatively the most to the risk of all other sectors during periods of distress affecting this sector. By contrast, the life insurance industry is the less risky financial sector in the United States for the same period of time. Furthermore, the life insurance industry appears to be highly interconnected with the non--life one, while diversified financial services seems to be highly interconnected with the banking system. This essentially means that during period of financial instability large losses in the banking sector affect the financial services sector more than the insurance sector.\newline

\textbf{Acknowledgments} This work has been partially supported by the 2011 Sapienza University of Rome Research Project.

\newpage
\appendix
\section{Tables}
\label{sec:appendix_A}
%

%
\begin{table}[!ht]
\captionsetup{font={small}, labelfont=sc}
%
\begin{small}
\resizebox{\columnwidth}{!}{%
\centering
 \smallskip
  \begin{tabular}{lcccccccc}\\
  \hline\hline
   Name & Min & Max & Mean$\times10^3$ & Std. Dev. & Skewness & Kurtosis & 1\% Str. Lev. & JB\\
    \hline
Banks 		& -0.318  & 0.377  & 0.930  & 0.042  & -0.050  & 19.786  & -0.106  & 13325.840  	\\
Fin. srvs		& -0.243  & 0.242  & 1.588  & 0.037  & -0.008  & 8.326  & -0.096  & 1341.471  		\\
Life insur. 		& -0.379  & 0.336  & 1.566  & 0.043  & -0.657  & 24.121  & -0.124  & 21178.483  	\\
Non--life Insur. & -0.272  & 0.157  & 1.076  & 0.030  & -0.661  & 12.363  & -0.085  & 4228.914  	\\
      \hline\hline
\end{tabular}}
\caption{Summary statistics of the US sector indexes form January, 2nd 1992 till June, 28th 2013. The eight column, denoted by ``1\% Str. Lev.'' is the 1\% empirical quantile of the returns distribution, while the last column, denoted by ``JB'' is the value of the Jarque-Ber\'a test-statistics.}
\label{tab:USSectors_data_summary_stat}
\end{small}
%
%
\end{table}
%

%
\begin{table}[!ht]
\captionsetup{font={small}, labelfont=sc}
%
\begin{small}
\centering
 \smallskip
  \begin{tabular}{cccc}\\
  \hline\hline
   L & log-likelihood & AIC & BIC \\
\cmidrule(lr){1-1} \cmidrule(lr){2-2} \cmidrule(lr){3-3} \cmidrule(lr){4-4}    
2  & 11889.544  & -23713.088  & -23410.697  \\
3  & 11713.089  & -23320.177  & -23053.355  \\
4  & 11969.138  & \textbf{-23788.276}  & \textbf{-23546.954}  \\
5  & 11946.912  & -23695.824  & -23197.420  \\
6  & 11973.013  & -23696.025  & -23066.727  \\
\hline\hline      
\end{tabular}
\caption{Log-likelihood, AIC and BIC values for Student-t Markov Switching models with different components. Bold faces indicates the selected model.}
\label{tab:DataSet_Model_Selection}
\end{small}
%
%
\end{table}
%

%
%
\begin{table}[!ht]
\captionsetup{font={small}, labelfont=sc}
%
\begin{small}
\centering
 \smallskip
  \begin{tabular}{cccccc}\\
  \hline\hline
 \multicolumn{1}{c}{\multirow{2}{*}{$\boldsymbol{\nu}$}}& State 1 & State 2 &State 3 & State 4\\
 \cmidrule(lr){2-2} \cmidrule(lr){3-3} \cmidrule(lr){4-4} \cmidrule(lr){5-5}
 & 15.6839 & 10.2542 & 11.0300 & 9.9473 \\
\hline
$\mathbf{Q}$& State 1 & State 2 &State 3 & State 4\\
\cmidrule(lr){1-1} \cmidrule(lr){2-2} \cmidrule(lr){3-3} \cmidrule(lr){4-4} \cmidrule(lr){5-5}
State 1& 0.8934 & 0.1066 & 0.0000 & 0.0000 \\
State 2& 0.0244 & 0.9608 & 0.0071 & 0.0077 \\
State 3& 0.0000 & 0.0000 & 0.9919 & 0.0081 \\
State 4& 0.0000 & 0.0052 & 0.0022 & 0.9926 \\
\hline\hline
\end{tabular}\\
\caption{ML estimates of the initial probability $\boldsymbol{\delta}$ and transition probability matrix $\mathbf{Q}$ of the Markov chain for the selected Student--t Markov Switching model with four components.}
\label{tab:DataSet_International_Estimates_2}
\end{small}
%
%
\end{table}
%

%
\begin{table}[!ht]
\captionsetup{font={small}, labelfont=sc}
%
\begin{small}
\centering
 \smallskip
  \begin{tabular}{lcccccc}\\
  \hline\hline
  $\boldsymbol{\mu}\times 10^3$&Banks& Fin. srvs & Life insur. & Non--life insur. \\
\hline
State 1& -6.8905 & -0.2288 & -8.1268 & -2.7188 \\
State 2& -4.5542 & -4.2744 & -1.9875 & -3.4971 \\
State 3& 1.0459 & 1.8574 & 1.2758 & 1.9089 \\
State 4& 3.5818 & 4.4010 & 4.3724 & 2.9295 \\
\hline
$\boldsymbol{\Lambda}\times 10^3$& Banks& Fin. srvs & Life insur. & Non--life insur. \\
\hline
State 1& 13.8891 & 7.0203 & 15.6855 & 4.3998 \\
State 1& 1.4302 & 1.6766 & 1.0333 & 1.0232 \\
State 1& 0.9998 & 0.6871 & 1.2204 & 0.3585 \\
State 1& 0.3427 & 0.5197 & 0.2827 & 0.3137 \\
\hline
$\boldsymbol{\Omega}_1$ & Banks& Fin. srvs & Life insur. & Non--life insur. \\
\hline
 Banks			& 1.0000 & & & \\
 Fin. srvs 			& 0.8399 & 1.0000 & & \\
 Life insur. 		& 0.8511 & 0.8248 & 1.0000 \\
 Non--life insur. 		& 0.6966 & 0.8013 & 0.8015 & 1.0000 \\
\hline
$\boldsymbol{\Omega}_2$ & Banks& Fin. srvs & Life insur. & Non--life insur. \\
\hline
Banks			& 1.0000 & & & \\
 Fin. srvs 			& 0.8573 & 1.0000 & & \\
 Life insur. 		& 0.7654 & 0.7771 & 1.0000 \\
 Non--life insur. 		& 0.7683 & 0.7881 & 0.8205 & 1.0000 \\
\hline
$\boldsymbol{\Omega}_3$ & Banks& Fin. srvs & Life insur. & Non--life insur. \\
\hline
Banks			& 1.0000 & & & \\
 Fin. srvs 			& 0.8807 & 1.0000 & & \\
 Life insur. 		& 0.8537 & 0.8939 & 1.0000 \\
 Non--life insur. 		& 0.7827 & 0.8472 & 0.8555 & 1.0000 \\
\hline
$\boldsymbol{\Omega}_4$ & Banks& Fin. srvs & Life insur. & Non--life insur. \\
\hline
Banks			& 1.0000 & & & \\
 Fin. srvs 			& 0.8620 & 1.0000 & & \\
 Life insur. 		& 0.7337 & 0.7678 & 1.0000 \\
 Non--life insur. 		& 0.7250 & 0.7622 & 0.7792 & 1.0000 \\
\hline\hline
\end{tabular}\\
\caption{ML parameter estimates of the selected Student--t Markov Switching model with four components where $\boldsymbol{\mu}$ are locations while the diagonal matrix $\bLambda$ and the full matrix $\bOmega$ are such that $\bSigma=\bLambda\bOmega\bLambda$.}
\label{tab:DataSet_International_Estimates_1}
\end{small}
%
%
\end{table}
%

%
\begin{sidewaystable}[h!]
\begin{center}
\begin{tabular}{c|cccc}
\hline\hline
&banks&fin. srvs&life Ins.&non--life Ins.\\
\midrule
banks &&\raisebox{-.5\height}{\includegraphics[width=0.20\linewidth, height=0.1\linewidth]{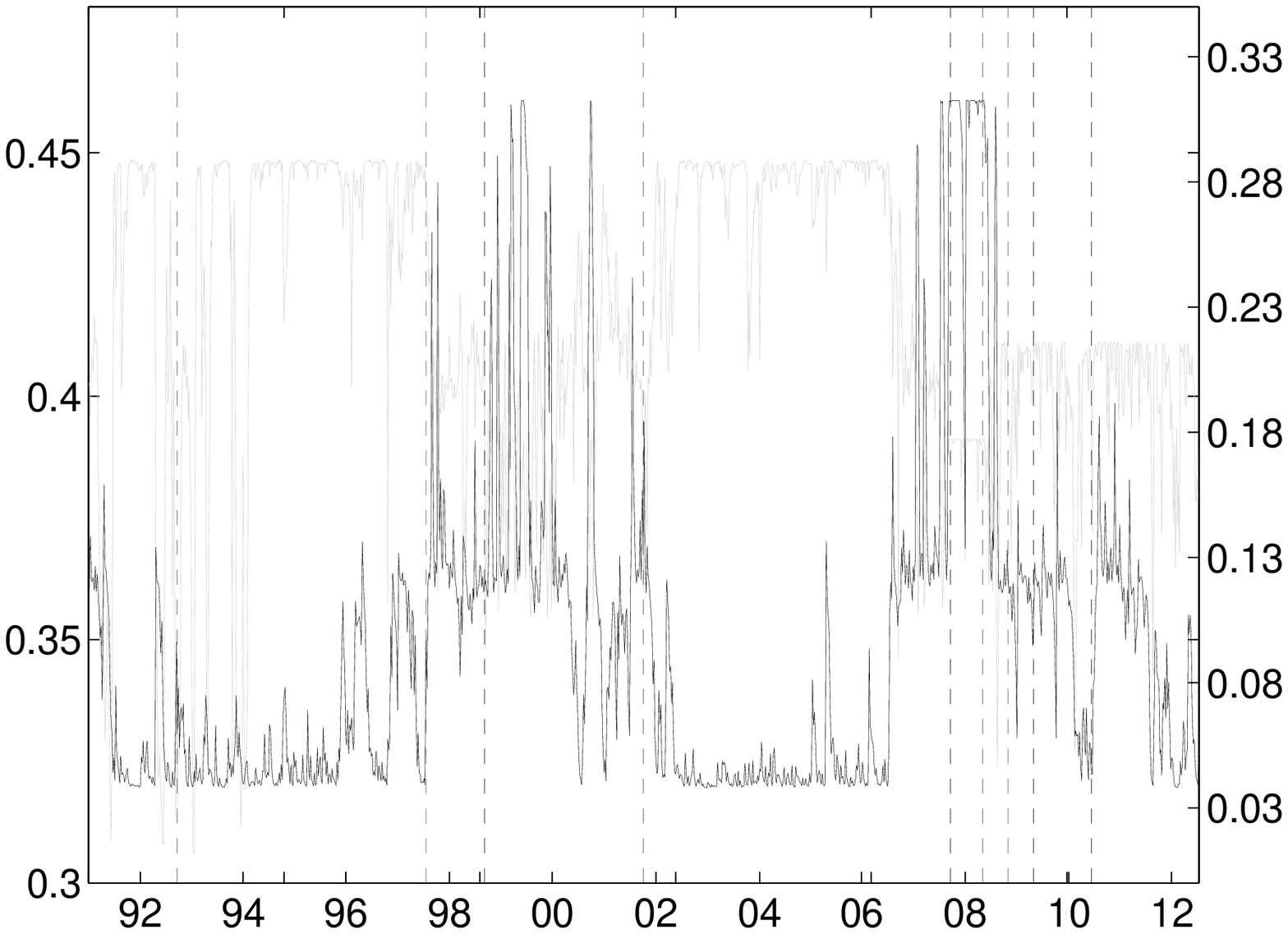}} &
\raisebox{-.5\height}{\includegraphics[width=0.20\linewidth, height=0.1\linewidth]{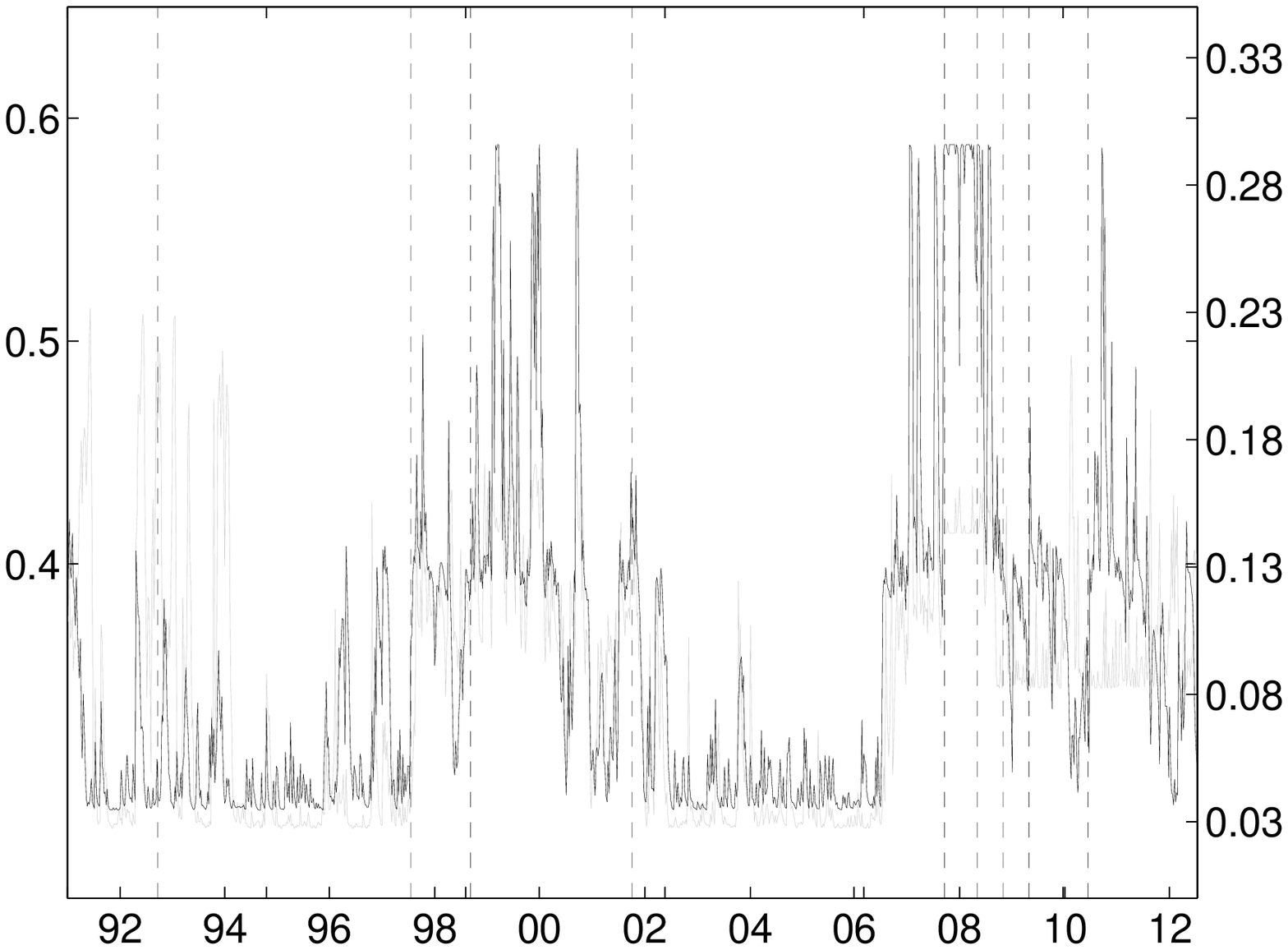}} &
\raisebox{-.5\height}{\includegraphics[width=0.20\linewidth, height=0.1\linewidth]{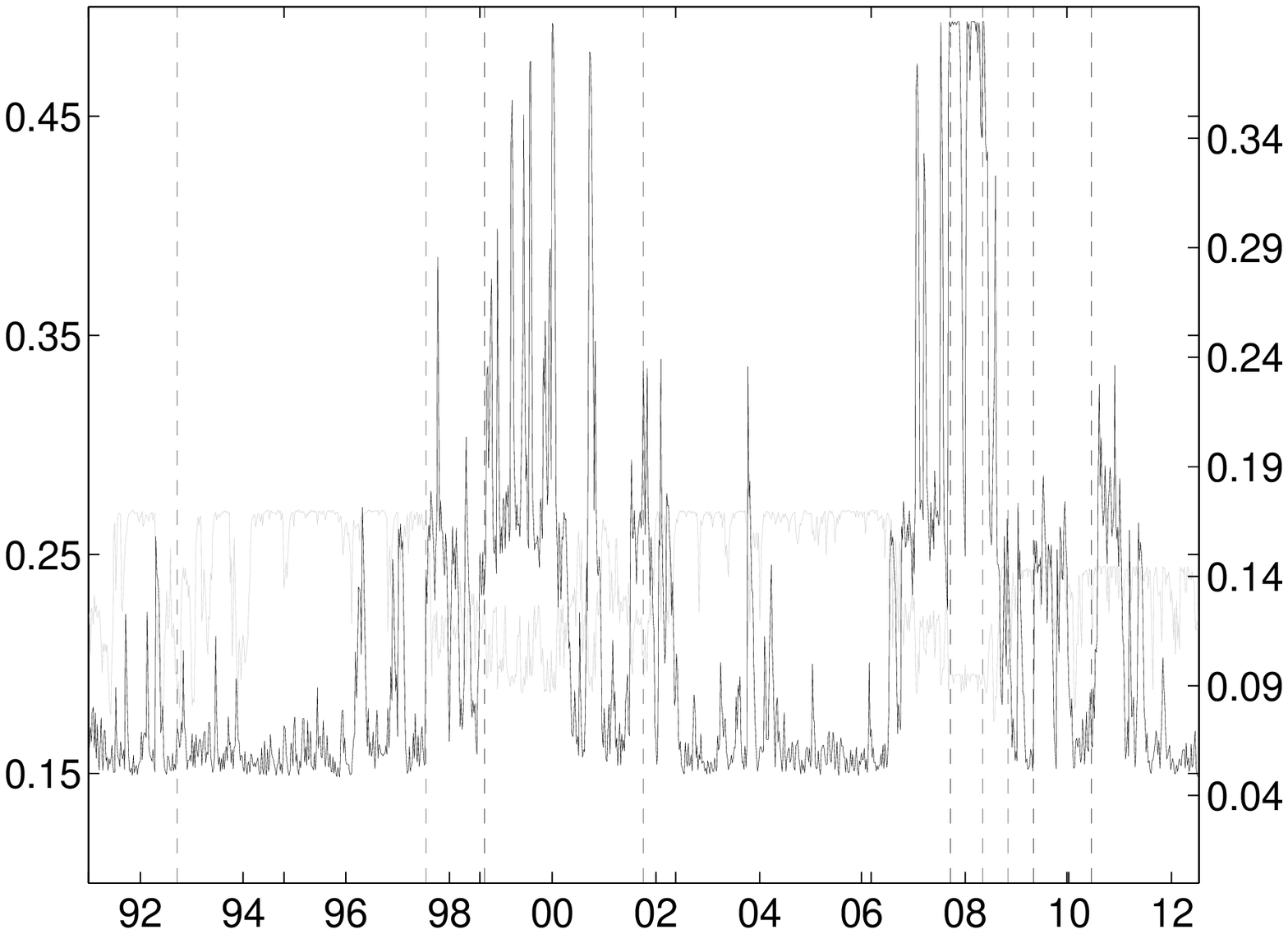}} \\[1cm]
fin. srvs  & \raisebox{-.5\height}{\includegraphics[width=0.20\linewidth, height=0.1\linewidth]{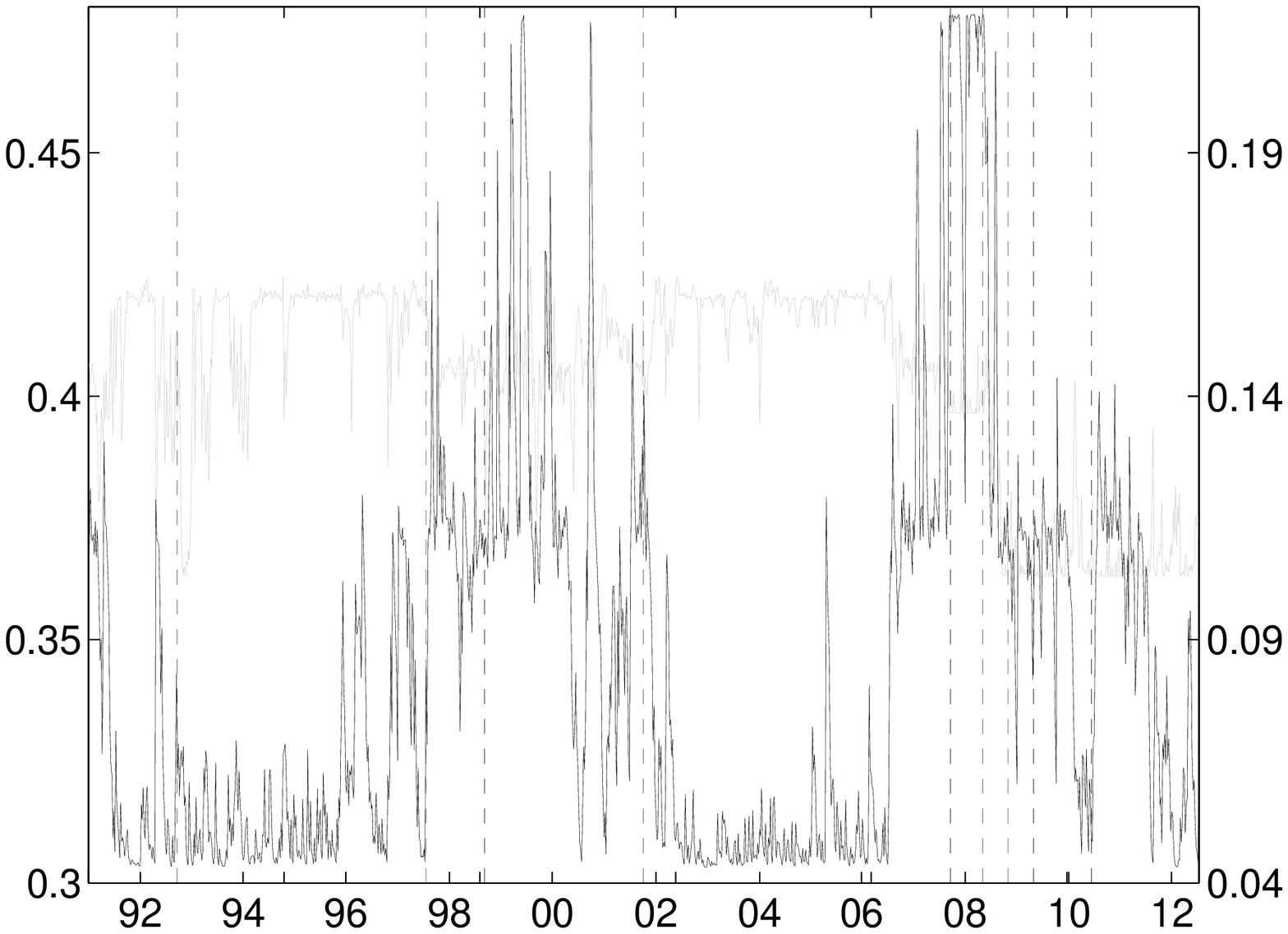}} & &
\raisebox{-.5\height}{\includegraphics[width=0.20\linewidth, height=0.1\linewidth]{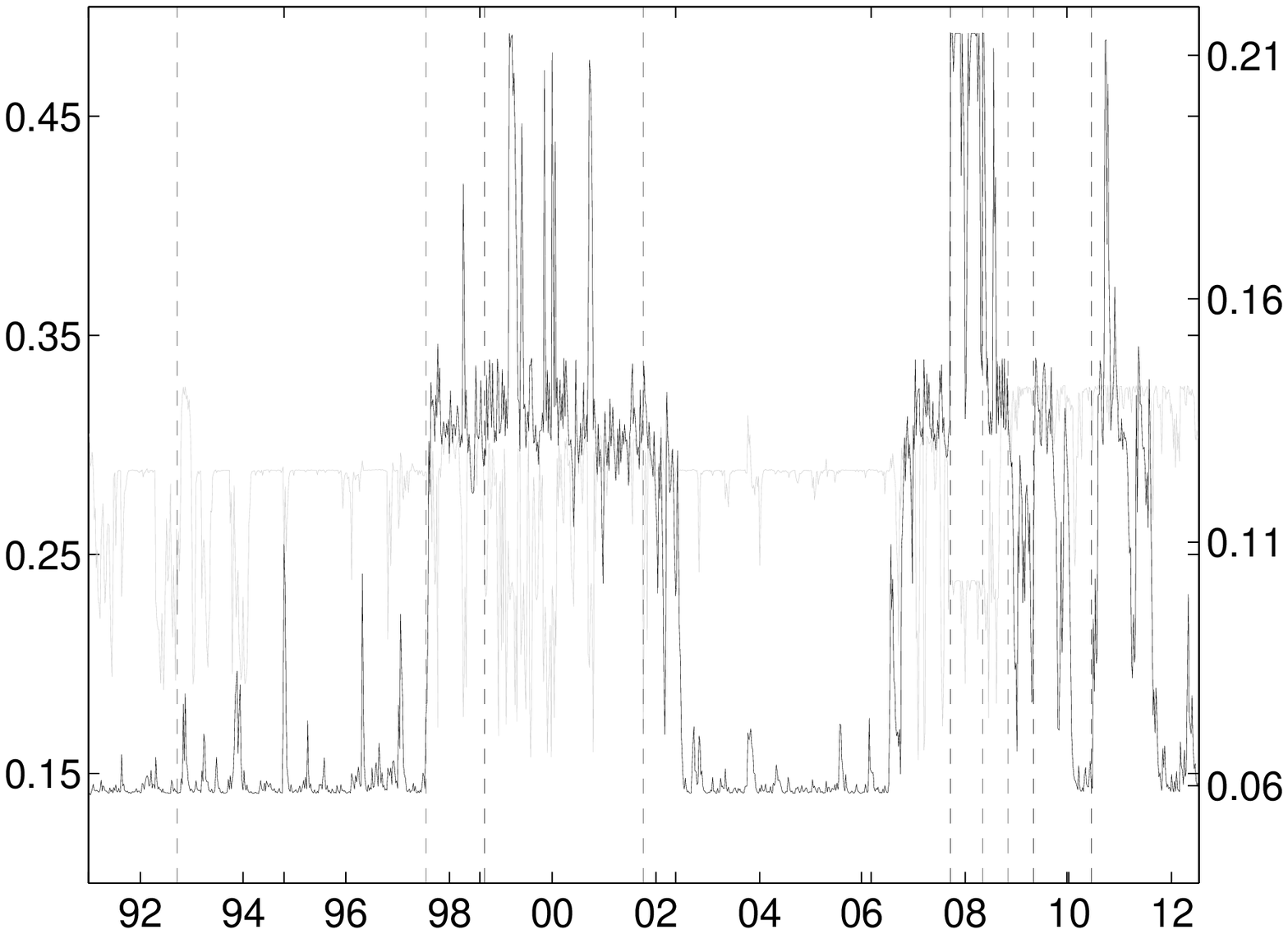}} & 
\raisebox{-.5\height}{\includegraphics[width=0.20\linewidth, height=0.1\linewidth]{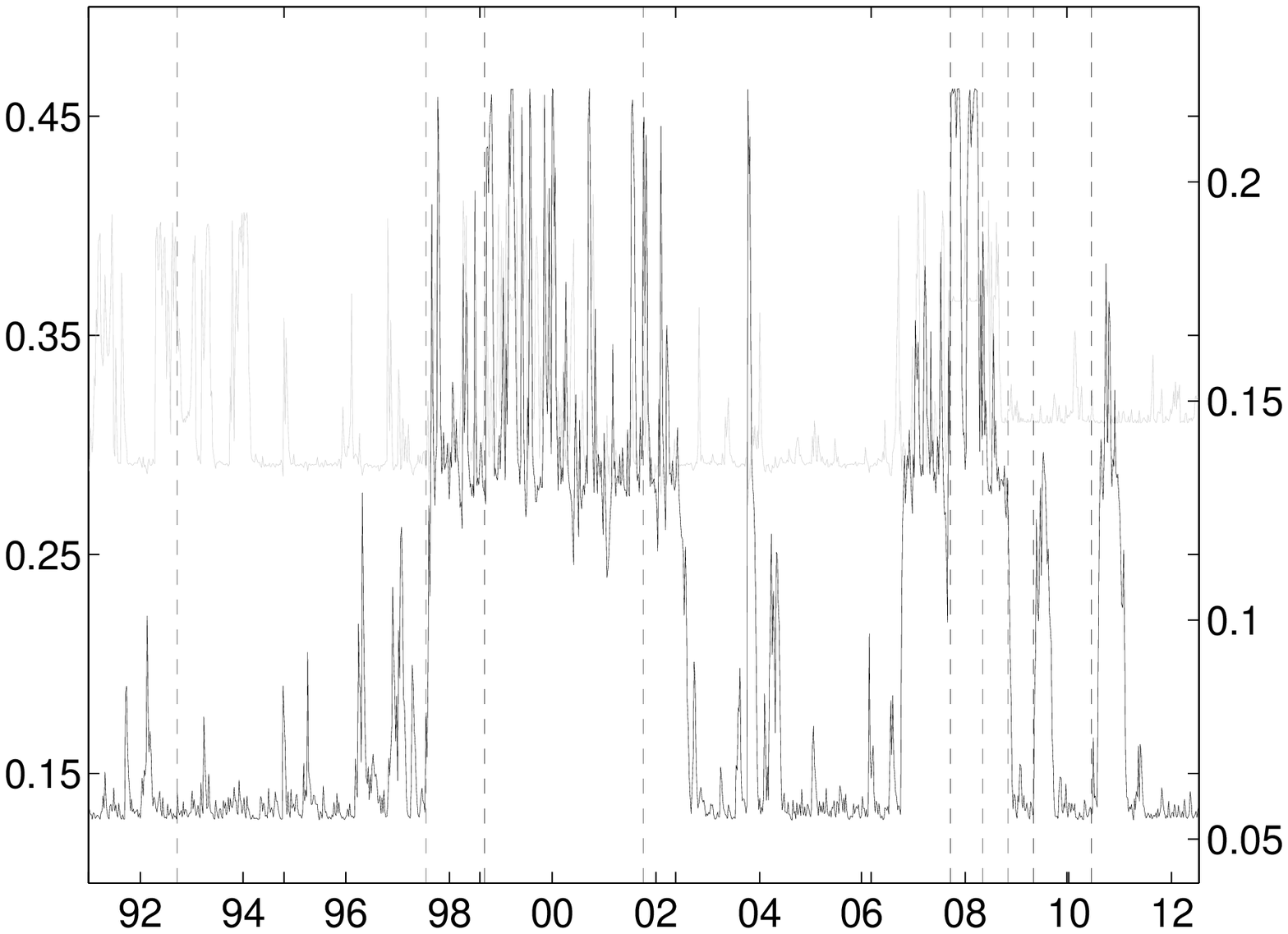}} \\[1cm]
life Ins. & \raisebox{-.5\height}{\includegraphics[width=0.20\linewidth, height=0.1\linewidth]{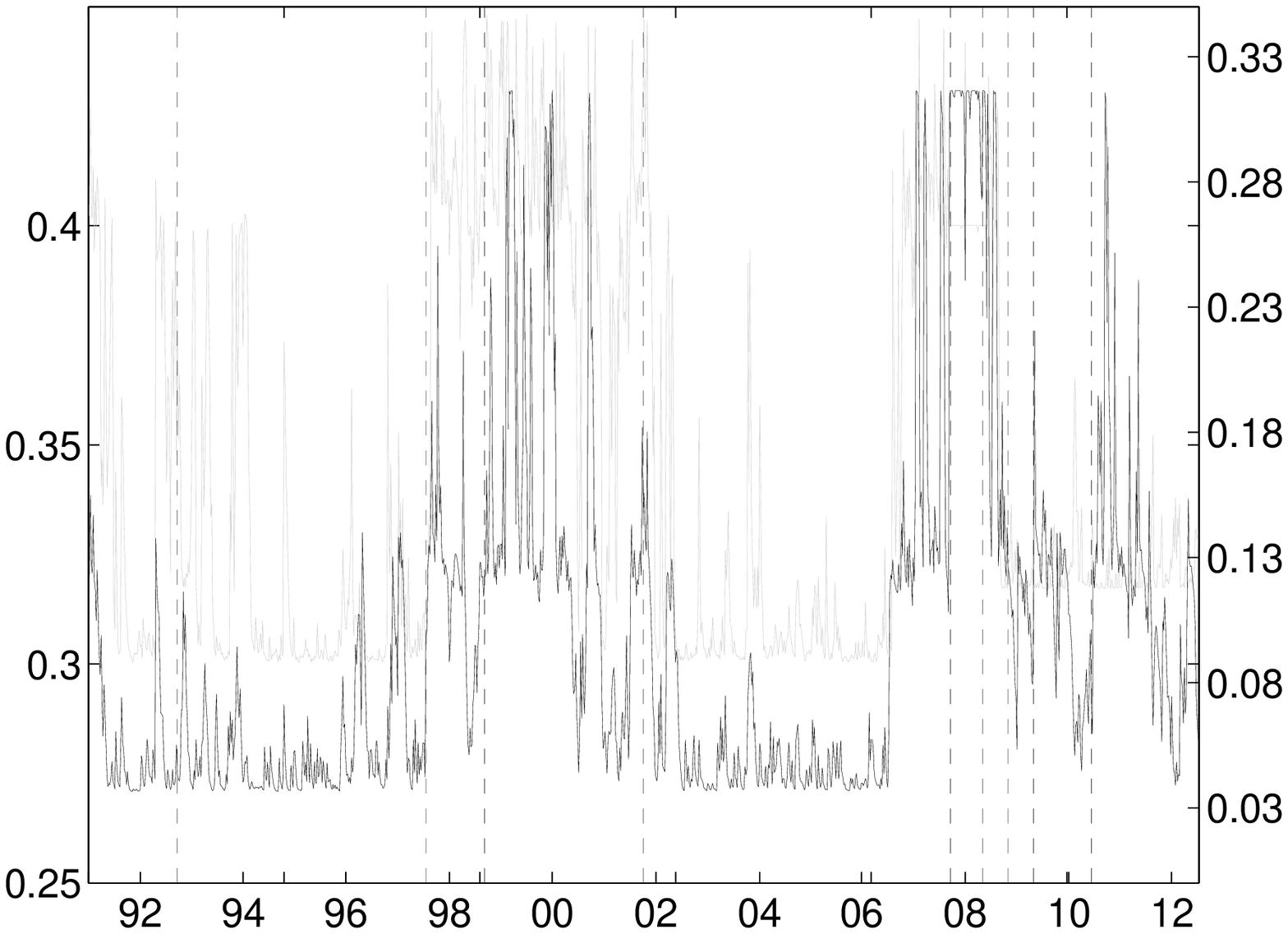}} &
\raisebox{-.5\height}{\includegraphics[width=0.20\linewidth, height=0.1\linewidth]{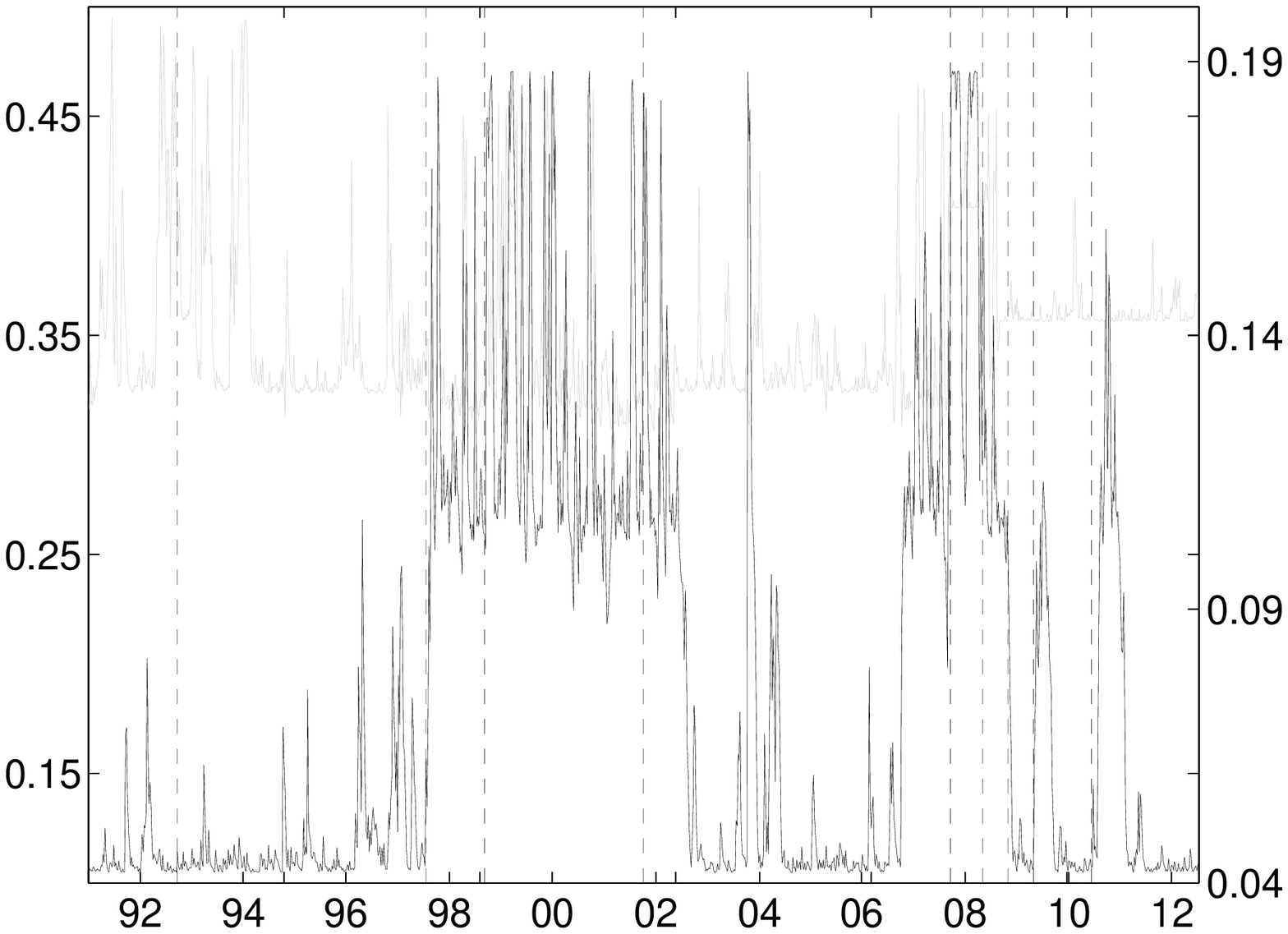}} & &
\raisebox{-.5\height}{\includegraphics[width=0.20\linewidth, height=0.1\linewidth]{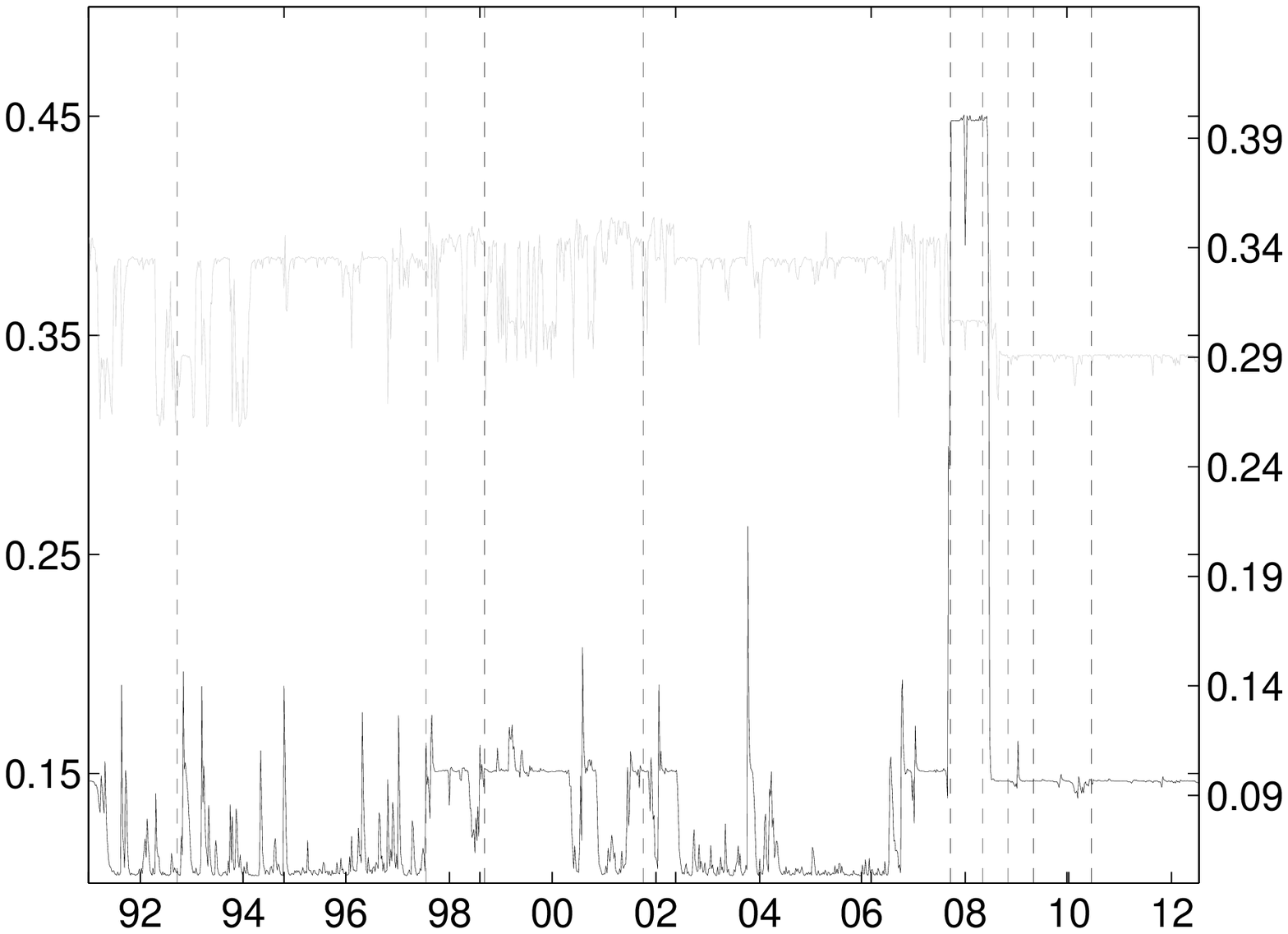}}\\[1cm]
non--life Ins. &\raisebox{-.5\height}{\includegraphics[width=0.20\linewidth, height=0.1\linewidth]{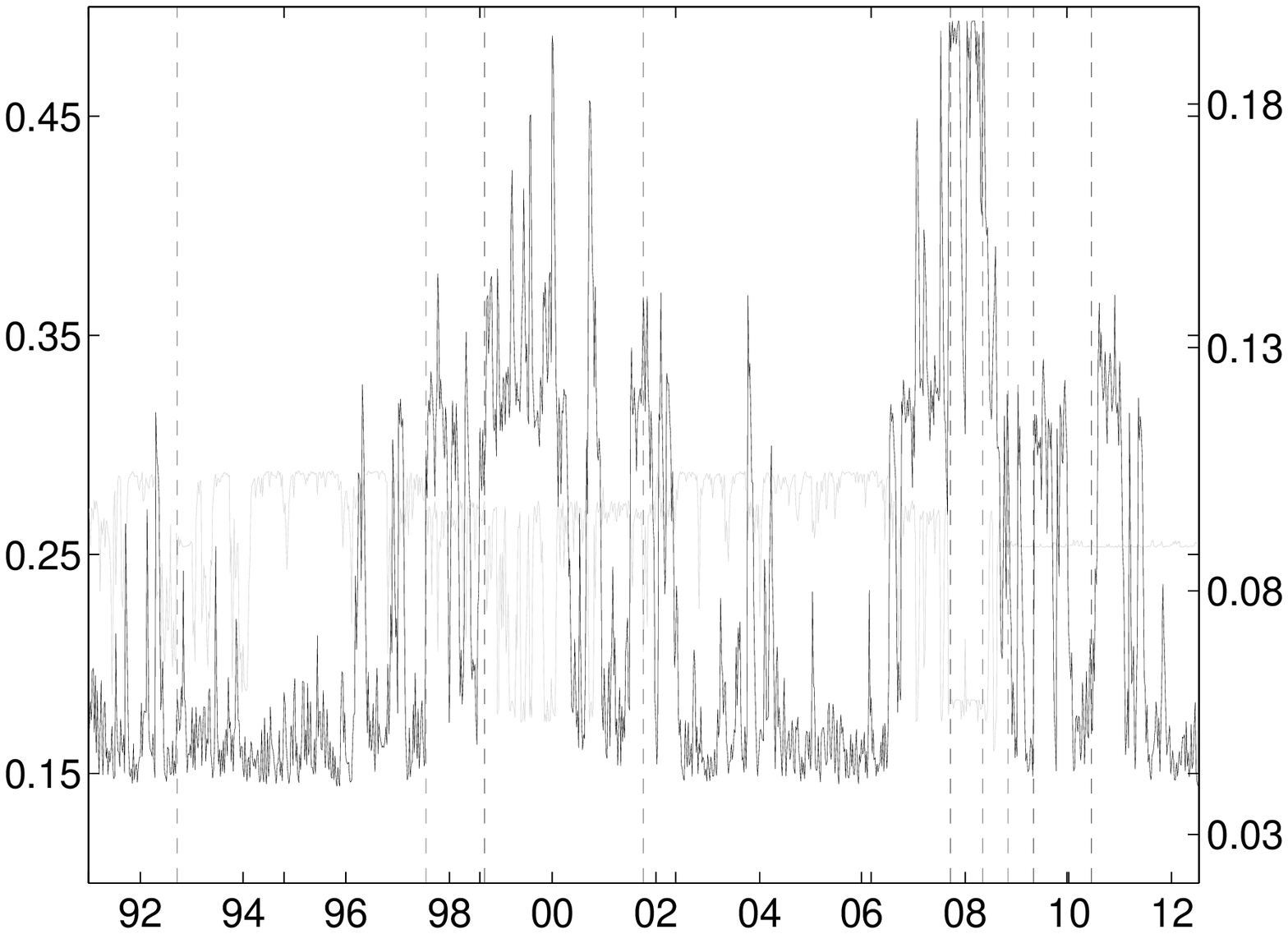}} &
\raisebox{-.5\height}{\includegraphics[width=0.20\linewidth, height=0.1\linewidth]{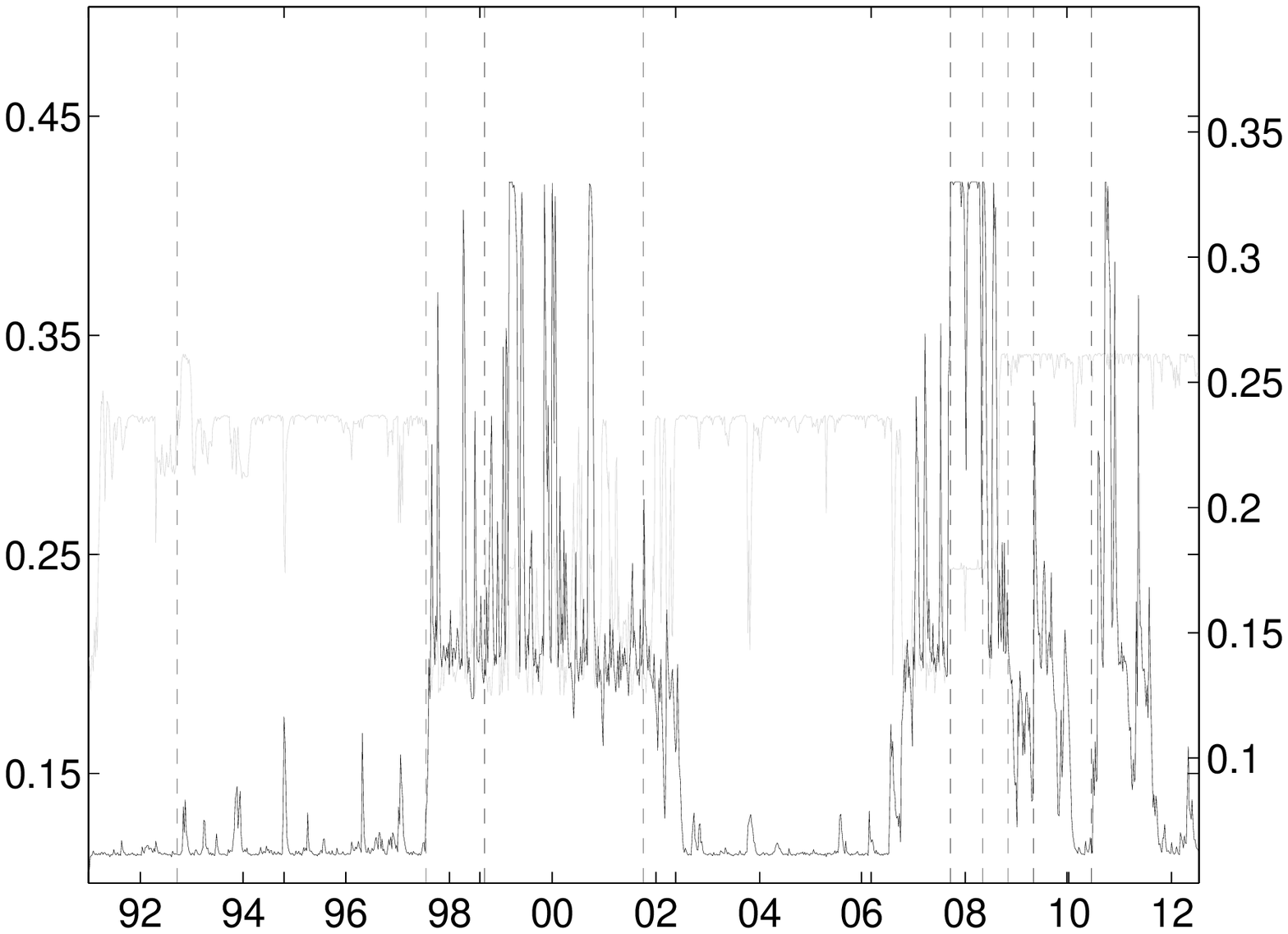}} &
\raisebox{-.5\height}{\includegraphics[width=0.20\linewidth, height=0.1\linewidth]{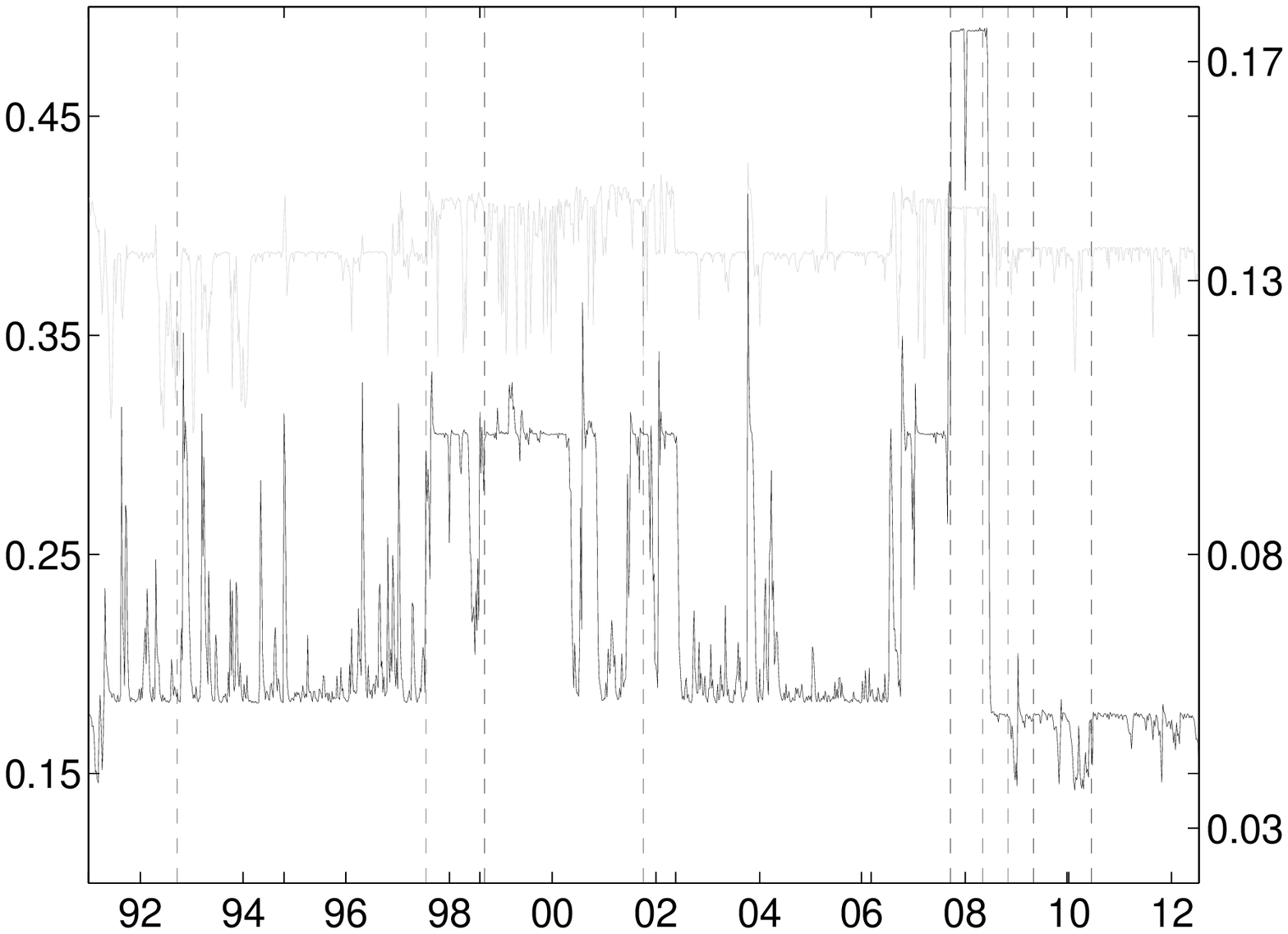}}\\[1cm]
\hline\hline
\end{tabular}
\captionsetup{font={small}, labelfont=sc}
\caption{\small{Comparison of the Shapley value $\Delta^\sM$CoES \textit{(light gray)} and the Adrian and Brunnermeier's standard $\Delta^\sM$CoVaR approach \textit{(light gray)} for all the sector against each other.
Vertical dotted lines represent major financial downturns: the ``Black Wednesday'' (September 16, 1992), the Asian crisis (July, 1997), the Russian crisis (August, 1998), the September 11, 2001 shock, the Bear Stearns hedge funds collapse (August 5, 2007), the Bear Stearns acquisition by JP Morgan Chase, (March 16, 2008), the Lehman's failure (September 15, 2008), the peak of the onset of the recent global financial crisis (March 9, 2009) and the European sovereign-debt crisis of April 2010 (April 23, 2010, Greek crisis).}}
\label{fig:USSector_Shapley_value_CoES}
\end{center}
\end{sidewaystable}
%

\clearpage
\newpage



\begin{thebibliography}{00}
\bibliographystyle{plainnat}
%
\bibitem[Adams, Z., F\"uss, R. and Gropp, R., (2011)]{adams_etal.2011} \textsc{Adams, Z., F\"uss, R. and Gropp, R., (2011).} Modeling spillover effects among financial institutions: A State--Dependent Sensitivity Value--at--Risk Approach. \textit{Working Paper EBS}.
\bibitem[Adrian, T. and Brunnermeier, M.K., (2011)]{adrian_brunnermeier.2011} \textsc{Adrian, T. and Brunnermeier, M.K., (2011).} \newblock CoVaR. \newblock{\em Working Paper Federal Reserve Bank of New York}. 
\bibitem[Amisano, G. and Geweke, J., (2011)]{amisano_geweke.2011} \textsc{Amisano, G. and Geweke, J., (2011).} \newblock Hierarchical Markov Normal Mixture models with applications to financial asset returns. \newblock{\em Journal of Applied Econometrics}, 26, pp. 1--29.
\bibitem[Bell, M. and Keller, B., (2009)]{bell_keller.2009} \textsc{Bell, M. and Keller, B., (2009).} \newblock Insurance and stability: the reform of insurance regulation \newblock{\em Zurich financial services Group}, Zurich, Switzerland.
\bibitem[Bernal, O., Gnabo, J.--Y. and Guilmin, (2013)]{bernal_etal.2013} \textsc{Bernal, O., Gnabo, J.--Y. and Guilmin, (2013).} Assessing the contribution of banks, insurance and other financial services to systemic risk. \newblock{\em Working paper SSRN}.
\bibitem[Bernardi, M., Gayraud, G. and Petrella, L., (2013a)]{bernardi_etal.2013a} \textsc{Bernardi, M., Gayraud, G. and Petrella, L., (2013).} \newblock Bayesian inference for CoVaR. Preprint arXiv:{\tt 1306.2834} [stat.ME], \url{http://arxiv.org/abs/1306.2834}.
\bibitem[Bernardi, M., Maruotti, A. and Petrella, L., (2013b)]{bernardi_etal.2013b} \textsc{Bernardi, M., Maruotti, A. and Petrella, L., (2013).} \newblock Multivariate Markov Switching models for tail risk interdependence. Preprint arXiv:{\tt 1312.6407} [stat.ME], \url{http://arxiv.org/abs/1312.6407}.
\bibitem[Billio, M., Getmansky, M., Lo, A.W. and Pellizon, L., (2012)]{billio_etal.2012} \textsc{Billio, M., Getmansky, M., Lo, A.W. and Pellizon, L., (2012).} Econometric measures of connectedness and systemic risk in the finance and insurance sectors. \textit{Journal of financial Economics}, 104, pp. 535--559.
\bibitem[Brechmann, E.C., Hendrich, K. and Czado, C., (2013)]{brechmann_etal.2013} \textsc{Brechmann, E.C., Hendrich, K. and Czado, C., (2013).} \newblock Conditional copula simulations for systemic risk stress testing. \newblock{\em Insurance: Mathematics and Economics}, 53, pp. 722--732.
\bibitem[Bulla, J., (2011)]{bulla.2011} \textsc{Bulla, J., (2011).} \newblock Hidden Markov Models with t Components. Increased Persistence and Other Aspects. \newblock{\em Quantitative Finance}, 11, pp. 459--475.
\bibitem[Cao, Z., (2013)]{cao.2013} \textsc{Cao, Z., (2013).} \newblock Multi--CoVaR and Shapley value: a systemic risk measure. \newblock{\em Banque de France Working paper}.
\bibitem[Capp\'e, O., Moulines, E. and Ryd\'en, T., (2005)]{cappe_etal.2005} \textsc{Capp\'e, O., Moulines, E. and Ryd\'en, T., (2005).} \newblock{\em Inference in Hidden Markov Models}. Springer Series in Statistics, Springer--Verlag, Berlin.
\bibitem[Castro, C. and Ferrari, S., (2014)]{castro_ferrari.2014} \textsc{Castro, C. and Ferrari, S., (2013).} \newblock Measuring and testing for the systemically important financial institutions. Forthcoming in \newblock{\em Journal of Empirical Finance}.
\bibitem[Chen, H., Cummins, J.D., Viswanathan, K.S., Weiss, M.A., (2013)]{chen_etal.2013} \textsc{Chen, H., Cummins, J.D., Viswanathan, K.S., Weiss, M.A., (2013).} \newblock Systemic risk and the interconnectedness between banks and insurers: an econometric analysis.\newblock{\em The Journal of Risk and insurance}, DOI: 10.1111/j.1539-6975.2012.01503.x.
\bibitem[Cont, R. and Moussa, A., (2010)]{cont_moussa.2010} \textsc{Cont, R. and Moussa, A., (2010)}. \newblock Too interconnected to fail: contagion and systemic risk in financial networks.\newblock{\em Financial Engineering Report 2010-03, Columbia University}.
\bibitem[Cont, R., Moussa, A. and Santos, E.B., (2012)]{cont_etal.2012} \textsc{Cont, R., Moussa, A. and Santos, E.B., (2012)}. \newblock Network structure and systemic risk in banking systems. \newblock{\em SSRN Working paper}.
\bibitem[Cummins, J.D. and M.A. Weiss, (2012)]{cummins_weiss.2012} \textsc{Cummins, J.D. and M.A. Weiss, (2012)}. \newblock Systemic Risk and the insurance Industry. Forthcoming in Georges Dionne, ed., \newblock{\em Handbook of insurance}, 2nd ed., Springer.
\bibitem[Dempster, A.P., Laird, N.M., Rubin, D.B., (1977)]{dempster_etal.1977} \textsc{Dempster, A.P., Laird, N.M., Rubin, D.B., (1977).} \newblock Maximum likelihood from incomplete data using the EM algorithm (with discussion). \newblock{\em Journal of the Royal Statistical Society, (series B)}, 39, pp. 1--39.
\bibitem[Dymarski, P., (2011)]{dymarski.2011} \textsc{Dymarski, P., (2011).} \newblock{\em Hidden Markov Models, Theory and Applications}, \newblock Rijeka, HR, Intech, pp. 207--222.
\bibitem[Geweke, J. and Amisano, G., (2010)]{geweke_amisano.2010} \textsc{Geweke, J. and Amisano, G., (2010).} \newblock Comparing and evaluating Bayesian predictive distributions of asset returns. \newblock{\em International Journal of Forecasting}, 26, pp. 216--230. 
\bibitem[Girardi, G. and Erg\"un, A.T., (2013)]{girardi_ergun.2013} \textsc{Girardi, G. and Erg\"un, A.T., (2013).} \newblock Systemic risk measurement: multivariate GARCH estimation of CoVaR. \newblock{\em Journal of Banking and Finance}, 37, pp. 3169--3180. 
\bibitem[Harrington, S., (2009)]{harrington.2009} \textsc{Harrington, S., (2009).} \newblock The financial crisis, systemic risk, and the future of insurance regulation. \newblock{\em Journal of Risk and insurance}, 76, pp. 785--819.
\bibitem[Jaeger--Ambrozewicz, M., (2012)]{jaeger-ambrozewicz.2012} \textsc{Jaeger--Ambrozewicz, M., (2012).} \newblock Closed form solutions of measures of systemic risk. \newblock{\em Journal of Risk and insurance}, Preprint arXiv:{\tt 1211.4173} [stat.ME], \url{http://arxiv.org/abs/1211.4173}.
\bibitem[Markose, S., Giansante S., and Shaghaghi, R.A., (2012)]{markose_etal.2012} \textsc{Markose, S., Giansante S., and Shaghaghi, R.A., (2012).} \newblock ``Too interconnected to fail'' financial network of U.S. CDS Market: topological fragility and systemic risk, \newblock{\em Journal of Economic Behavior and Organization}, 83, pp. 627--646.
\bibitem[Podlich, N. and Wedow, M., (2013)]{Podlich_etal.2013}\textsc{Podlich, N. and Wedow, M., (2013).} \newblock Are insurers SIFIs? A MGARCH model to measure interconnectedness. \newblock{\em Applied Economics Letters}, 20, pp. 677--681.
\bibitem[Ryd\'en, T., (2008)]{ryden.2008} \textsc{Ryd\'en, T., (2008).} \newblock EM versus Markov chain Monte Carlo for estimation of Hidden Markov models: a computational perspective. \newblock{\em Bayesian Analysis}, 3, pp. 659--688.
\bibitem[Shapley, L., (1953)]{shapley.1953} \textsc{Shapley, L., (1953).} \newblock A value for n--person Games, \newblock{\em Annals of Mathematical Studies}, 28, pp. 307--317.
\bibitem[Tarashev, N., Borio, C. and Tsatsaronis, K., (2010)]{tarashev_etal.2010} \textsc{Tarashev, N., Borio, C. and Tsatsaronis, K., (2010).} \newblock Attributing systemic risk to individual institutions: methodology and policy applications, \newblock{\em BIS working parper} No. 308.
\bibitem[Zucchini, W. and MacDonald, I.,
(2009)]{zucchini_macdonald.2009} \textsc{Zucchini, W. and MacDonald, I.,
(2009).} \newblock{\em Hidden Markov Models for Time Series: an Introduction using R}, \newblock Chapman \& Hall/CRC.
%
\end{thebibliography}
\end{document}